\documentclass[longauth]{aa}  

\usepackage{graphicx}
\usepackage{txfonts}
\usepackage[hidelinks,colorlinks=true,linkcolor=blue,citecolor=blue]{hyperref}
\usepackage[dvipsnames]{xcolor}
\usepackage{tabularx}
\newcolumntype{L}[1]{>{\raggedright\arraybackslash}p{#1}}
\newcolumntype{C}[1]{>{\centering\arraybackslash}p{#1}}
\newcolumntype{R}[1]{>{\raggedleft\arraybackslash}p{#1}}
\usepackage{multicol}
\usepackage{multirow}
\usepackage{makecell}
\usepackage{cancel}
\newcommand{\hi}{H{\sc i}}

\begin{document}

   \title{The BINGO project VIII:} 
   \subtitle{On the recoverability of the BAO signal on \hi\ intensity mapping simulations}
    
    \titlerunning{The BINGO project VIII: On the recoverability of the BAO signal on \hi\ IM simulations}

   \author{Camila P. Novaes\fnmsep\thanks{camila.novaes@inpe.br}\inst{1}
          \and
          Jiajun Zhang\inst{2}
          \and
          Eduardo J. de Mericia\inst{1}
          \and
          Filipe B. Abdalla\inst{1,3,4,5}
          \and
          Vincenzo Liccardo\inst{1}
          \and
          Carlos A. Wuensche\inst{1}
          \and
          Jacques Delabrouille\inst{6,7,8,9}
          \and
          Mathieu Remazeilles\inst{10,11}
          \and
          Larissa Santos\inst{12,13}
          \and
          Ricardo G. Landim\inst{14}
          \and
          Elcio Abdalla\inst{4}
          \and
          Luciano Barosi\inst{15}
          \and 
          Amilcar Queiroz\inst{15}
          \and
          Thyrso Villela\inst{1,16,17}
          \and 
          Bin Wang\inst{12,13}
          \and
          Francisco A. Brito\inst{15}
          \and 
          Andr\'e A. Costa\inst{12,18}
          \and 
          Elisa G. M. Ferreira\inst{4,19}
          \and
          Alessandro Marins\inst{4}
          \and 
          Marcelo V. dos Santos\inst{4,15}
          }

   \institute{
   Instituto Nacional de Pesquisas Espaciais, Av. dos Astronautas 1758, Jardim da Granja, S\~ao Jos\'e dos Campos, SP, Brazil\\
              \email{camilapnovaes@gmail.com}
    \and
    Shanghai Astronomical Observatory, Chinese Academy of Sciences, Shanghai 200030, China
    \and
    University College London, Gower Street, London, WC1E 6BT, UK
    \and
    Instituto de F\'{i}sica, Universidade de S\~ao Paulo, R. do Mat\~ao, 1371 - Butant\~a, 05508-09 - S\~ao Paulo, SP, Brazil
    \and
    Department of Physics and Electronics, Rhodes University, PO Box 94, Grahamstown, 6140, South Africa
    \and
    CNRS-UCB International Research Laboratory, Centre Pierre Bin\'etruy, IRL2007, CPB-IN2P3, Berkeley, USA
    \and
    Laboratoire Astroparticule et Cosmologie (APC), CNRS/IN2P3, Universit\'e Paris Diderot, 75205 Paris Cedex 13, France 
    \and
    IRFU, CEA, Universit\'e Paris-Saclay, 91191 Gif-sur-Yvette, France 
    \and
    Department of Astronomy, School of Physical Sciences, University of Science and Technology of China, Hefei, Anhui 230026
    \and
    Instituto de Física de Cantabria (CSIC-Universidad de Cantabria), Avda. de los Castros s/n, E-39005 Santander, Spain
    \and
    Jodrell Bank Centre for Astrophysics, Department of Physics and Astronomy, The University of Manchester, Oxford Road, Manchester, M13 9PL, UK
    \and
    Center for Gravitation and Cosmology, College of Physical Science and Technology, Yangzhou University, 225009, China
    \and 
    School of Aeronautics and Astronautics, Shanghai Jiao Tong University, Shanghai 200240, China 
    \and
    Technische Universit\"at M\"unchen, Physik-Department T70, James-Franck-Strasse 1, 85748, Garching, Germany
    \and
    Unidade Acad\^emica de F\'{i}sica, Universidade Federal de Campina Grande, R. Apr\'{i}gio Veloso, Bodocong\'o, 58429-900 - Campina Grande, PB, Brazil 
    \and
    Centro de Gest\~ao e Estudos Estrat\'egicos SCS Qd 9, Lote C, Torre C S/N Salas 401 a 405, 70308-200 - Bras\'ilia, DF, Brazil
    \and
    Instituto de F\'{i}sica, Universidade de Bras\'{i}lia, Campus Universit\'ario Darcy Ribeiro, 70910-900 - Bras\'{i}lia, DF, Brazil 
    \and
    College of Science, Nanjing University of Aeronautics and Astronautics, Nanjing 211106, China
    \and
    Kavli IPMU (WPI), UTIAS, The University of Tokyo, 5-1-5 Kashiwanoha, Kashiwa, Chiba 277-8583, Japan
             }

   \date{Received ; accepted }

 
  \abstract
   { A new and promising technique for observing the Universe and study the dark sector is the intensity mapping of the redshifted 21\,cm line of neutral hydrogen (\hi). 
   The \textbf{B}aryon Acoustic Oscillations [BAO] from \textbf{I}ntegrated \textbf{N}eutral \textbf{G}as \textbf{O}bservations (\textbf{BINGO}) radio telescope will use the 21\,cm line to map the Universe in the redshift range  $0.127 \le z \le 0.449$, in a tomographic approach, with the main goal of probing BAO.
   }
   {This work presents the forecasts of measuring the transversal BAO signal during the BINGO Phase 1 operation.}
   {We use two clustering estimators: the two-point angular correlation function (ACF), in configuration space, and the angular power spectrum (APS), in harmonic space, and a template-based method to model the ACF and APS estimated from simulations of the BINGO region and extract the BAO information. 
   The tomographic approach allows the combination of redshift bins to improve the template fitting performance. 
   We compute the ACF and the APS for each of the 30 redshift bins and measure the BAO signal in 3 consecutive redshift blocks (lower, intermediate and higher) of 10 channels each. Robustness tests are used to evaluate several aspects of the BAO fitting pipeline for both clustering estimators.  
   }
   {We find that each clustering estimator shows different sensitivities to specific redshift ranges, although both of them perform better at higher redshifts. 
   In general, the APS estimator provides slightly better estimates, with smaller uncertainties and larger probability of detection of the BAO signal, achieving $\gtrsim 90$\% at higher redshifts.
   We investigate the contribution from instrumental noise and residual foreground signals and find that the former has the greater impact, getting more significant as the redshift increases, in particular the APS estimator. 
   Indeed, including noise in the analysis increases the uncertainty up to a factor of $\sim 2.2$ at higher redshifts. 
   Foreground residuals, in contrast, do not significantly affect our final uncertainties. 
   }
   {In summary, our results show that, even including semi-realistic systematic effects, BINGO has the potential to successfully measure the BAO scale in radio frequencies.}


   \maketitle
%

\section{Introduction} \label{sec:introduction}

Recent cosmological observations, such as distance measurements with type Ia supernovae, observations of cosmic microwave background (CMB) temperature and polarization anisotropies, and the detection of baryon acoustic oscillations (BAO) in large scale structure spectroscopic surveys, have been key in establishing the current standard cosmological model, $\Lambda$CDM. 

Although a remarkable fit to existing observations, $\Lambda$CDM requires 95\% of the energy content of the universe to be in the form of dark matter ($\sim$ 25\%), and a puzzling component responsible for the current accelerated expansion of the universe, dark energy, with approximately 70\%. Confidence in the model requires the identification of these two components of the Universe \citep{Abdalla:2020ypg}.
Otherwise the $\Lambda$CDM  model will be subject to being challenged by alternative interpretations, many of which involving modifications of our theory of gravitation on large scales.

The BAO feature, appearing as a fixed scale in both the CMB and large scale structure (LSS) data (a geometrical probe), is originated in the early Universe, when the temperature was high enough to keep photons and baryons coupled.
The BAO are the imprints left by acoustic waves traveling at relativistic speed,
generated by the gravitational infall of baryons (and dark matter) into the potential wells of dark matter balanced by the radiation pressure pushing out these baryons from overdense regions. 
The comoving distance they traveled until recombination when the baryons were released from the drag of the photons defines the BAO scale \citep{2003/dodelson-book}. 
This characteristic scale is then imprinted in the LSS distribution and evolves with structure formation as a standard ruler that can be measured as a function of the redshift \citep{2007/eisenstein-seo-white, 2007/seo-eisenstein}.
This makes the BAO one of the most powerful and well established tool to investigate the history of the acceleration of the expansion of the Universe \citep{2013/weinberg}. 

The first statistically significant measurements of the BAO feature imprinted in the galaxy distribution were done by \cite{2005/eisenstein} analyzing the Sloan Digital Sky Survey (SDSS) data, and by \cite{2005/cole}, using the 2dF Galaxy Survey (2dFGS).  
After that, several other detections of the BAO scale have been performed using different biased tracers of the dark matter in several redshift ranges. 
Among them we can cite the analyses of SDSS data at various stages of data gathering \citep{2012/anderson-sdss, 2014/anderson-sdss, 2017/alam-sdss, 2018/ata-boss-collab, 2021/alam-sdss}, besides the WiggleZ Dark Energy Survey (DES) \citep{2016/hinton} and the 6dFGS \citep{2018/carter-6df} analyses. 
More recently,  data releases of the DES have been analyzed for the detection of the BAO signature through different clustering measurement methods: in configuration, in harmonic and Fourier spaces, and for 2- or 3-dimensional distribution of sources \citep{2019/camacho, 2019/abbott-des, 2021/abbott-des}. 
These results have used the common template-based fitting method to study the BAO feature from clustering estimates. 
Alternative approaches for measuring the BAO scale in the angular (2-dimensional) 2- and 3-point statistics use an empirical parametric fit technique, for example, \cite{2011/sanchez, 2012/carnero, 2013/de-simoni, 2016/carvalho, 2017/carvalho, 2018/deCarvalho, 2020/deCarvalho} and \cite{2021/deCarvalho}.

A new and very promising way of measuring BAO is through the detection of structures as traced by the redshifted diffuse 21\,cm hyperfine transition line of neutral hydrogen (\hi) through the so called intensity mapping (IM) technique \citep{2008/chang}. 
Compared to radio surveys using the 21\,cm emission to detect individual galaxies, limited by the low luminosity of the 21\,cm line emission, the IM technique can cover a larger volume of the Universe in a much shorter time, using instruments with relatively small, and consequently cheaper, collecting area, as discussed and evaluated in \cite{2013/battye} and \cite{2021/abdalla_BINGO-project} \citep[see also,][]{2008/chang, 2008/loeb-wyithe}. 
The idea is to measure the overall \hi\ brightness temperature field, similarly to what is done for CMB temperature fluctuations, but mapping the Universe as a function of redshift. 
This is possible given the high abundance of hydrogen, so that a hydrogen map is expected to be a powerful tracer of the underlying total matter content of the Universe. 
In this sense, through redshift surveys in radio frequencies, the 21\,cm IM constitutes a new window for observing the Universe, and for studying the dark sector.

To exploit this new observational window, several radio instruments, already observing or still under construction, will survey a large volume of the Universe through the 21\,cm IM technique and measure the BAO feature. 
Among them we can mention the Square Kilometer Array\footnote{\url{https://www.skatelescope.org/}} \citep[SKA;][]{2020/ska}; MeerKAT\footnote{\url{https://www.ska.ac.za/science-engineering/meerkat}} \citep{2017/santos}, a precursor of the SKA; the Five-Hundred-Meter Aperture Spherical Radio Telescope, currently the largest single dish telescope in the world \cite[FAST;][]{2011/fast}; the Canadian Hydrogen Intensity Mapping Experiment\footnote{\url{https://chime-experiment.ca/}} \citep[CHIME;][]{2014/chime}; Tianlai\footnote{\url{http://tianlai.bao.ac.cn/wiki/index.php/Main_Page}} \citep{2012/tianlai}; HIRAX \citep{2022/hirax}; and the Baryon Acoustic Oscillations from Integrated Neutral Gas Observations\footnote{\url{https://www.bingotelescope.org/en/}}  \citep[BINGO;][]{2013/battye, 2015/bigot-sazy, 2021/abdalla_BINGO-project}, described in the next section. 

In this work we evaluate what to expect from future BINGO observations in terms of measuring the transversal (angular) BAO signal. To perform this task we employ and compare two clustering estimators, the two-point angular correlation function (ACF), in configuration space, and the angular power spectrum (APS), in harmonic space. 
To our knowledge, this is the first forecasting study of the BAO detection from 21\,cm IM simulations using ACF. 
The anisotropic two-point correlation function estimator $\xi(r_\perp,r_\parallel)$, was explored by \cite{2021/avila} for the case of a SKA-like survey employing simulations, and by \cite{2021/kennedy}, for a theoretical modeling approach focusing in MeerKAT survey. 
Both of them focus on the impact of the instrumental beam smoothing and foreground removal in recovering the BAO scale. 
\cite{2017/villaescusa} also explore the SKA case, assessing the BAO detection through the radial power spectrum, under the presence of instrumental effects and foreground contamination, and the impact of the angular resolution imposed by a single-dish instrument. 
They find that the telescope beam will compromise the detection of the isotropic BAO feature at redshifts $z \gtrsim 1$, while the radial BAO seems to be robust against the foreground removal. 
These published results confirm the potential of future 21\,cm IM observations for detecting the BAO signature. 

In the present work, we use a template-based method to model the APS and ACF estimated from two types of mock realizations, mimicking future BINGO observations, and extract the BAO information from each of them. 
We use the covariance matrices calculated from the mocks to construct the likelihood corresponding to each template and, using a maximum likelihood estimator, we estimate the parameters of the model for each simulation.
We apply this template fitting procedure over three sets of consecutive bins, so that we can evaluate the BAO detection at three different redshift intervals.

This paper is organized as follows. 
Section \ref{sec:instrument} briefly describes the main aspects of the BINGO project. 
Section \ref{sec:theory} introduces the \hi\ clustering theory and the template model constructed for each estimator. 
Section \ref{sec:simulations} presents the details about the cosmological 21\,cm simulations, the characteristics of the instrumental noise and the foreground contamination, as well as the foreground cleaning process employed here. 
Section \ref{sec:methodology} summarizes our methodology for the clustering measurements and the BAO fitting process, and the covariance matrix construction. 
The results from all our analyses are discussed in Section \ref{sec:results}, while the conclusions are summarized in Section \ref{sec:conclusions}. 

\section{The BINGO telescope} \label{sec:instrument}

BINGO is a radio telescope in construction in a site (see \cite{Peel:2019} for the site selection process)  located in Para\'iba State, northeastern Brazil (latitude: 7$^\circ$ 2' 27.6'' S; longitude: 38$^\circ$ 16' 4.8'' W; altitude: 350 to 460\,m). Its main scientific goal is to measure the BAO signal imprinted in the 21\,cm distribution in the redshift range $0.127 \le z \le 0.449$ \citep{2021/abdalla_BINGO-project}. 
This goal will be achieved through an IM survey covering 5324 squares degrees, using the telescope in a sky transit mode, with a $14.75^{\circ}$-wide declination strip centered at a declination $\delta=-15^{\circ}.$ 

The telescope has a crossed-Dragone design, with a 40-m diameter primary paraboloid and a 34-m diameter secondary hyperboloid. During BINGO Phase 1, the telescope will operate with a focal plane containing 28 horns \citep{2021/wuensche_BINGO-instrument, 2021/abdalla_BINGO-optical_design}. 
Each horn is sensitive to circular polarization and is coupled to a correlation receiver with 4 amplifier chains, and should perform at an expected system temperature $T_{sys} = 70$\,K. 
The optical design will produce a suitable angular resolution for the BAO signal, with a full width half maximum $\theta_{\mbox{\sc fwhm}} = 40'$ of the beam in the central frequency $\nu = 1120$ MHz.
In a second phase, we intend for BINGO to operate with 28 additional horns, totaling 56 horns.

The redshift range covered by BINGO corresponds to the frequency interval $980 - 1260$ MHz, which is quite complementary to instruments like  CHIME \citep[$400 - 800$ MHz;][]{2014/chime}. 
A more detailed description of the BINGO project, its scientific goals and instrument status, are available in previous publications of the collaboration \citep{2021/abdalla_BINGO-project,2021/wuensche_BINGO-instrument}.

\section{Modeling the BAO signal} \label{sec:theory}
\subsection{\hi\ clustering}

The intensity mapping of the 21\,cm signal measures the brightness temperature of the average \hi\ intensity in a given volume of the Universe, which can be written as function of the redshift, $z$, as 
\citep[see, e.g.,][]{2013/battye,2013/hall}%
\begin{equation}
    \bar{T}_{\rm HI}(z) = 188 \, h\, \Omega_{\rm HI}(z) \frac{(1+z)^2}{E(z)} \, {\rm mK}\, ,
\end{equation}
where $H_0 = 100\, h \,\, {\rm km \,\, s}^{-1} {\rm Mpc}^{-1}$ is the Hubble constant, $E(z) = H(z)/H_0$, and $\Omega_{\rm HI}$ is the \hi\ density parameter. 
Fluctuations of the \hi\ brightness temperature, $\delta T_{\rm HI}$, are a biased tracer of the dark matter fluctuations, $\delta$. Their Fourier transform can be written, in terms of the growth function $D(z)$, as %
\begin{equation}
    \delta T_{\rm HI}(\mathbf{k}, z) = D(z) \, \bar{T}_{\rm HI}(z) \, b_{\rm HI}(z) \, \delta(\mathbf{k}, 0) \, ,
\end{equation}
where $b_{\rm HI}(z)$ is the bias factor as a function of redshift $z$, and $\delta(\mathbf{k}, 0)$ is the underlying dark matter distribution at $z=0$. 
In this way, the \hi\ power spectrum is given by $P_{\rm HI} \approx [ D(z) \, \bar{T}_{\rm HI}(z) \, b_{\rm HI}(z) ]^2 P(k)$, where $P(k)$ is the matter power spectrum at $z=0$. 

As detailed by \cite{2013/battye} and \cite{2016/seehars}, since the 21\,cm IM surveys the Universe within tomographic bins of redshift, one can project the 3D quantity $\delta T_{\rm HI}(\chi(z)\hat{n}) = \bar{T}_{\rm HI}(z) \delta_{\rm HI}(\chi(z)\hat{n})$ on the sky by integrating it along the line of sight $\hat{n}$ as
\begin{equation}
    \delta T_{\rm HI}(\hat{{n}}) = \int dz \, \phi(z) \, \bar{T}_{\rm HI}(z) \, \delta_{\rm HI}(\chi(z) \, \hat{n}) \, ,
\end{equation}
where $\delta_{\rm HI}(\chi(z) \, \hat{n})$ is the density fluctuation of neutral hydrogen at the comoving distance to redshift $z$, $\chi(z)$, and $\phi(z)$ is the projection kernel\footnote{Here, we assume a top-hat kernel, that is, $\phi(z) = 1/(z_{max}-z_{min})$ for $z_{min} < z < z_{max}$ and $\phi(z) = 0$ outside the redshift bin. } (a window function of observation). 
Decomposing the \hi\ temperature fluctuations in spherical harmonics, we have \citep[see also,][]{2011/sobreira, 2021/costa_BINGO-forecast}
\begin{equation} \label{eq:sph_harm}
     \delta T_{\rm HI}(\hat{n}) = \sum_{\ell = 0}^{\infty} \sum_{m = -\ell}^{\ell} a_{\ell m} Y_{\ell m}(\hat{n}) \, ,
\end{equation}
with the $a_{\ell m}$ harmonic coefficients written as
\begin{equation}
    a_{\ell m} = 4 \pi i^\ell \int dz \, \phi(z) \, \bar{T}_{\rm HI}(z) \, \int \frac{d^3k}{(2\pi)^3}   \, \delta_{\rm HI}(\vec{\rm k},z) \, j_\ell(k \chi(z)) \, Y^*_{\ell m}(\vec{\rm k}) \, , 
\end{equation}
where $j_\ell$ is the spherical Bessel function. 
Therefore, from the above equations, we can define the APS of the temperature fluctuations, $C_\ell = \langle \, |a_{\ell m}|^2 \rangle$, as a function of the matter power spectrum, such that \citep[see also][for the analogous case of a galaxy distribution]{2019/loureiro}
\begin{equation} \label{eq:Cls}
    C^{ij}_\ell = \frac{2}{\pi} \int dk \, W^i_{{\rm HI},\ell}(k) \, W^j_{{\rm HI},\ell}(k) \, k^2 \, P(k) \, ,
\end{equation}
where the indices $i$ and $j$ denote two tomographic bins. For $i = j$ and $i \neq j$, we obtain auto- and cross-APS respectively \cite[see also,][]{2021/costa_BINGO-forecast}.
The redshift dependence is given by the window function
\begin{equation} \label{eq:Wl}
    W^i_{{\rm HI},\ell}(k) = \int  dz \, b_{\rm HI}(z) \, \phi(z) \, \bar{T}_{\rm HI}(z) \, D(z) \, j_\ell(k \chi(z)) \, .
\end{equation}

In a similar way to the APS one can also define the ACF, $\omega(\theta)$, defined as the probability to find galaxies separated by an angle $\theta$.
For a tomographic bin around $z$, the ACF can be defined as a function of the matter 3D spatial correlation function $\xi(r)$ \citep{2011/sobreira, 2011/crocce, 2011/sanchez, 2013/de-simoni}
\begin{eqnarray}
    \omega(\theta) & \equiv & \langle \delta T_{\rm HI}(\hat{{n}}) \, \delta T_{\rm HI}(\hat{n} + \vec{{\theta}}) \rangle \\ \label{eq:w_theta1}
    & = & \int dz_1 \, f(z_1) \int dz_2 \, f(z_2) \, \xi(r(z_1, z_2, \theta), \bar{z}) \, , \label{eq:w_theta2}
\end{eqnarray}
where $f(z) =  b_{\rm HI}(z) \phi(z) \bar{T}_{\rm HI}(z)$ and $r(z_1, z_2, \theta)$ is the radial comoving distance between spatial fluctuations $\delta_{\rm HI}$ at redshifts $z_1$ and $z_2$ separated by an angle $\theta$.  
This equation neglects the time evolution of the matter correlation function inside the bin, so that it can be evaluated at a given $\bar{z}$ (the average redshift of the bin) as
\begin{equation}
    \xi(r, \bar{z}) = \frac{1}{2\pi} \int dk k^2 j_0(r k) P(k,\bar{z}) \, ,
\end{equation}
where $j_0$ is the spherical Bessel function of zeroth order and $P(k,\bar{z})$ is the matter power spectrum at redshift $\bar{z}$. 

Alternatively, following \cite{2011/sobreira} and \cite{2011/crocce} and using Eq. (\ref{eq:sph_harm}), we can still obtain the ACF as
\begin{equation} \label{eq:w_theta3}
    \omega(\theta) = \bigg\langle \sum_{\ell = 0}^{\infty} \sum_{\ell = -\ell}^{\ell} a_{\ell m} a_{\ell' m'} Y_{\ell m}(\hat{n}) Y_{\ell' m'}(\hat{n} + \vec{\theta}) \bigg\rangle \, ,
\end{equation}
and, as a function of the APS, as
\begin{eqnarray} \label{eq:w_theta_from_Cls}
    \omega(\theta) &=& \sum_{\ell m} C_\ell Y_{\ell m}(\hat{n}) Y_{\ell m}(\hat{n} + \vec{\theta}) \label{eq:w_theta4}\\
    &=& \sum_{\ell} C_\ell \frac{2\ell + 1}{4\pi} P_\ell(\cos \theta) \, ,
\end{eqnarray}
where $P_\ell$ are the Legendre polynomials.

The theoretical auto- and cross-$C_\ell$'s employed in this paper, both as input to the 21\,cm log-normal simulations (see Section \ref{sec:mocks}) and to construct the BAO templates, are calculated using the Unified Cosmological Library for $C_\ell$'s ({\tt UCLCL}) code \citep{2019/loureiro, 2017/mcleod}. 
This code implements Eqs. (\ref{eq:Cls}) and (\ref{eq:Wl}) using the {\tt CLASS} Boltzmann Code \citep{2011/lesgourgues, 2011/blas} to estimate the primordial power spectra and transfer function. 
To account for all processes involved in the evolution of the Universe, we use the window function $W^{Tot,i}(k) = W^i_{{\rm HI},\ell}(k) + W^i_{RSD,\ell}(k)$. 
In addition to Eq. (\ref{eq:Wl}), it also includes a term describing the redshift space distortion (RSD) effect, $W^i_{RSD,\ell}(k)$.  
A detailed description of how the two terms are implemented in the {\tt UCLCl} code can be found in \cite{2019/loureiro}.

\subsection{BAO template}

The extraction of the BAO features from data clustering estimates is commonly performed by fitting a template model, derived from a parameterization of the matter power spectrum \cite[see, e.g.,][ and references in the Introduction]{2014/anderson-sdss}. 
Following the same approach, we write this parameterization as a function of the linear power spectrum, $P_{\rm lin}$, and the no-wiggle (no BAO feature) power spectrum, $P^{nw}$, 
\begin{equation} \label{eq:P_k}
    P^{temp}(k) = [ P^{lin}(k) - P^{nw}(k) ] \, e^{-k^2 \Sigma_{nl}^2} + P^{nw}(k) \, .
\end{equation}
We employ the {\tt nbodykit}\footnote{\url{https://github.com/bccp/nbodykit}} code to obtain both $P^{lin}(k)$ and $P^{nw}(k)$, using the transfer functions calculated by the {\tt CLASS} code and the analytic calculation by \cite{1998/eisenstein-hu}, respectively. 
The exponential term, $e^{-k^2 \Sigma_{nl}^2}$, takes into account the effect of non-linear structure growth by damping the signal around the BAO scale. 
The damping scale, $\Sigma^2_{\rm nl} = (\Sigma^2_{\perp} + \Sigma^2_{\parallel})/2$, is written in terms of the components along ($\Sigma_{\parallel}$) and across ($\Sigma_{\perp}$) the line of sight, taking into account the RSD effect. 
\cite{2007/seo-eisenstein} have shown that these components can be written as $\Sigma_{\perp} = 10.4 D(z) \sigma_8$ (prediction for real space)  and $\Sigma_{\parallel} = (1+f) \Sigma_{\perp}$, where $f$ is the growth rate of cosmic structures \cite[see also discussion by][]{2018/chan,2018/ata-boss-collab}. 
Since we already account for the RSD in the window function $W^i_{{\rm HI},\ell}$ when projecting $ P^{temp}(k)$ into the APS (Eqs. (\ref{eq:Cls}) and (\ref{eq:Wl})), we can consider $\Sigma_{\rm nl} = \Sigma_{\perp}$. 
For our fiducial cosmology (see Section \ref{sec:mocks}), we then fix the damping scale at $\Sigma_{\rm nl} = 7.6, 7.1$, and $6.7$ $h^{-1}$Mpc, appropriate values for the lower, intermediate and higher redshift intervals in which we split the BAO fitting analyses, as discussed later. Then, from the projection of $ P^{temp}(k)$ into $C^{temp}(\ell)$, we construct the template used for the APS analysis 
\begin{equation} \label{eq:APS-temp}
    C(\ell) = B \, C^{temp}(\ell/\alpha) + \sum_q A_q \, \ell^q \, ,
\end{equation}
where $\alpha$, $B$ and $A_q$ are free parameters, the last two of them intended to absorb linear and non-linear bias effects, noise, uncertainties in the RSD, besides any other difference between the data points and the full shape template, such as those introduced by systematic effects. 

For the ACF analyses, the template model is constructed substituting the same $C^{temp}(\ell)$ in Eq. (\ref{eq:w_theta_from_Cls}) to calculate $\omega^{temp}(\theta)$, so that we guarantee the consistency among both templates, obtaining
\begin{equation} \label{eq:ACF-temp}
    \omega(\theta) = B \omega^{temp}(\alpha \theta) + \sum_q \frac{A_q}{\theta^q} \, ,
\end{equation}
where $\alpha$, $B$ and $A_q$ are free parameters, as in Eq. (\ref{eq:APS-temp}).
The last term in both templates, appearing as functions of $\ell$ and $\theta$, can be written so that the number $q$ of $A_q$ parameters optimize our fitting pipeline. 
Section \ref{sec:robustness-tests} shows results from testing different degrees of freedom, that is, different choices for minimum and maximum values to run the $q$ index, for both $C_\ell$ and $\omega(\theta)$ templates.

In both cases the $\alpha$ parameter, the most important parameter of the analysis, is the so called shift parameter, associated to the change in the BAO peak position with respect to a fiducial cosmology, 
\begin{equation}
    \alpha = \frac{D_A(z)/r_d}{(D_A(z)/r_d)_{fid}} \, ,
\end{equation}
where $D_A$ is the angular diameter distance and $r_d$ is the sound horizon scale at the drag epoch. 
Then, $\alpha$ characterizes any observed deviation with respect to the model, so that $\alpha > 1$ ($\alpha < 1$) indicates a shift of the acoustic peak to smaller (larger) scales.
Since the fiducial cosmology adopted to model the BAO templates and to generate the synthetic data  are the same, we expect to find $\alpha \approx 1$ when fitting them to the APS and ACF clustering computed from the simulations, allowing to test the methodology and to predict what to expect from the BAO analysis of the future BINGO data.

\section{Synthetic data preparation} \label{sec:simulations}

In this section we describe the different sets of simulations employed in the analyses.
Sky maps are produced in the {\tt HEALPix} pixelization scheme \citep{2005/gorski}, with $N_{side} = 256$. 
In most of the tests we use a large set of (fast) log-normal simulations of the 21\,cm signal generated by the {\tt FLASK}\footnote{ \url{http://www.astro.iag.usp.br/~flask/}} code \citep{2016/xavier}, hereafter {\tt FLASK} mocks. 
We  also test the BAO fitting pipeline over a smaller data set based on the density contrast from N-body simulations, as described below, hereafter N-body mocks. 

Semi-realistic mock data sets (referred as BINGO-like simulations) include instrumental noise, beams effects, the BINGO sky coverage scan and simulated foreground signals that will contaminate 21\,cm observations in the BINGO  frequency range.
A foreground cleaning pipeline, outlined in Sec.~\ref{sec:gnilc}, is applied to a small set of BINGO-like simulations, constructed from a subset of the {\tt FLASK} mocks, to mitigate the effect of the foreground signal on the 21\,cm signal and estimate the residual contamination in the recovered maps. 

\subsection{BINGO-like simulations}

\subsubsection{Cosmological signal} \label{sec:mocks}

\begin{enumerate}
    \item[\it (a)] {\it Fast log-normal distributions}
\end{enumerate}
The first set of 21\,cm IM simulations consists of full-sky log-normal realizations generated with  {\tt FLASK}.
This is a publicly available code able to produce two- or three-dimensional tomographic realizations of an arbitrary number of random astrophysical fields and reproducing the desired cross-correlations between them. 
{\tt FLASK} takes as input the auto- and cross-APS, $C^{ij}(\ell)$, previously calculated for each of the $i$ and $j$ redshift slices and creates 2D {\tt HEALPix} maps of correlated log-normal realizations of the projected 21\,cm signal in each redshift bin (z-bin). 
We use the fiducial $C^{ij}(\ell)$ computed using the {\tt UCLCl} code. Cosmological parameters match those from WMAP 5-yr results \citep[Table 2, Five-Year Mean values, of][]{2009/wmap5}, $\Omega_m = 0.26$, $\Omega_b = 0.044$, $\Omega_{\Lambda} = 0.74$, and $H_0 = 72 \, {\rm km} \,{\rm s}^{-1} {\rm Mpc}^{-1}$, for consistency with the N-body mock simulations described later. 
Although more recent parameter constraints have been obtained with the \textit{Planck} satellite \citep{2018/planck-VI}, our pipeline and results does not drastically depend on the exact cosmological parameter values. 
Here we are not interested in constraining these parameters, but to show the consistency of our method and the detection of BAO. 

We generate a total of 1500 log-normal realizations, each of them corresponding to 30 \hi\ maps, one for each tomographic z-bin. 
Note that, despite of our \textit{ad-hoc} choice of 30 channels ($\delta \nu = 9.33$\,MHz width each), the BINGO hardware and data analysis pipeline allows different choices for the number of channels. 
Consequently, their width can be chosen according to what is best suitable for cosmological analyses, for example, as function of the efficacy of foreground cleaning procedure \citep{2022/mericia_BINGO-component_separation_II}. 
Tests on how the BAO measurement is impacted by using a different number of channels are left for future work. 

\begin{enumerate}
    \item[\it (b)] {\it N-body simulations}
\end{enumerate}

A second set of mocks is generated using the FoF (Friend-of-Friends) halo catalogs from HR4 N-body simulation \citep{kim2015horizon}. 
HR4 has a box size of 3150 Mpc/h and uses WMAP 5-yr cosmological parameters. 

First, we select 100 random observer locations in the simulation box, to mimic 100 different realizations in just one simulation.
We then construct the light cone catalog from snapshots at $z=0.1,0.15,0.2,0.3,0.4,$ and $0.5$. According to the designated frequency bin, the halos from the closest two snapshots are selected. These halos are further filtered randomly, and the selection probability is proportional to the radial distance of this bin and the corresponding comoving distance at the redshift of the snapshot. Therefore, the generated light cone halo catalog is a random mixture of the snapshots. We validate the method by comparing the angular power spectrum and halo mass function to the light cone halo catalog provided by HR4. Both results are consistent and meet our requirements.

We make the full sky 21\,cm brightness temperature map from the light cone halo catalog following the method described in \citet{2021/zhang-BINGO-mock_simulations}. 
By further considering redshift distortion effects, we construct 100 realizations of full sky mock maps, each containing 30 redshift bins.
Since BINGO exploits only a strip of the sky, we get independent realizations by randomly rotating the full sky maps before including realistic characteristics of the BINGO observations. 
Using 4 independent rotations\footnote{Spherical rotations as performed by the {\tt healpy} package.}, we generate another 400 mock maps in the BINGO footprint using the 100 original full sky maps. 
After performing these simulations, we have 1500 {\tt FLASK} and 500 N-body mocks for testing our BAO detection pipelines.

\subsubsection{Foreground signals, instrumental effects and sky coverage} \label{sec:foreg}

Following previous BINGO papers by \cite{2021/fornazier_BINGO-component_separation} and \cite{2021/liccardo_BINGO-sky-simulation}, our simulations include the contribution of seven foreground components, including Galactic synchrotron, free-free, thermal dust and anomalous microwave (AME) emissions, extragalactic thermal and kinetic Sunyaev-Zel'dovich (SZ) effects, and unresolved radio point sources, all generated using a recent version of the \textit{Planck} Sky Model software \citep[PSM;][]{2013/delabrouille}. 

The specific configuration of the PSM code that was used to generate our foreground emissions is as follows. Synchrotron is based on the 408 MHz all sky map produced by \cite{2015/remazeilles}, extrapolated to BINGO frequencies with a spatially variable spectral index map following the model derived by \cite{2008/miville-deschenes}. Free-free emission is simulated using a template given by the $H\alpha$ emission map from \cite{2003/dickinson}, with a uniform frequency scaling  over the sky and slowly varying with frequency.
For the AME, usually described as a radio emission produced by the rapid rotation of electric dipoles associated to small dust grains, we use a high resolution thermal dust template from \textit{Planck} observations, scaled to low frequency according to the ratio between AME and thermal dust as found by \cite{2016/planck-XXV}, and extrapolated it to BINGO frequencies using a single emission law. 
A template for the thermal dust emission is obtained by applying Generalized Needlet Internal Linear Combination\footnote{This is the component separation code also employed here for cleaning the 21\,cm maps; see Section \ref{sec:gnilc} for references and a brief description of the code.} ({\tt GNILC}) code to Planck 2015 data. Dust spectral index and temperature maps, used to extrapolate thermal dust emission across frequencies using a single modified blackbody in each pixel, are obtained from fits of GNILC dust maps at different frequencies. 
Although AME and thermal dust emission are subdominant at BINGO frequencies, both are taken into account here for completeness.

The extragalactic emission from thermal and kinetic SZ effects are modeled according to prior knowledge of existing galaxy clusters for a number density as a function of mass and redshift predicted by the cosmological model of interest, given by \cite{2008/tinker}. 
The unresolved radio point sources component is simulated using catalogs of observed sources from 850 MHz and 4.85 GHz. A population of sources below detection threshold is simulated on the basis of theoretical number counts.
A more complete discussion of foreground simulation for BINGO frequency can be found in \cite{2021/fornazier_BINGO-component_separation} and \cite{2021/liccardo_BINGO-sky-simulation}.


We also take into account the expected thermal (white) noise level per pixel, employing a system temperature of 70 K. 
This level is estimated considering 28 horns, each of them observing at a constant elevation, achieving a survey area of 5324 deg$^2$ (a sky fraction of $f_{sky} \sim 0.13$).
We assume 5 years of observation and the horn arrangement designed for the Phase 1 of observations, as discussed in previous BINGO publications \citep{2021/wuensche_BINGO-instrument, 2021/abdalla_BINGO-optical_design,2021/liccardo_BINGO-sky-simulation}.
For good sampling of the map in declination, the position of the horns will be shifted to displace their pointing on the sky by a fraction of a beam width in elevation. 
This will homogenize the observation time over the innermost BINGO area. 
A detailed description of the noise simulations employed here can be found in \cite{2021/fornazier_BINGO-component_separation}.

Note that, although this paper does not account for the low frequency ($1/f$) noise, we are aware it is an important complication in real data and can be very detrimental to {\hi} IM data if not efficiently removed \citep{2015/bigot-sazy,2021/li}. 
The $1/f$ noise is correlated across the frequency band and, when projecting the observed signal into a map, it appears as stripes, that is, as large scale spatial fluctuations \citep{2018/harper}.
In this sense, since such instrumental effect has the potential to impact our analyses at lowest redshifts, where the BAO scale is comparable to the $15^\circ$ stripe of the BINGO coverage, it is being addressed in various working groups across the collaboration and our findings will be presented in upcoming works of BINGO series. 
Moreover, it is also worth remembering that another instrumental effect relevant for intensity mapping experiments is the leakage of polarized foregrounds into the intensity data \citep{2015/alonso, 2021/cunnington}. 
Unlike the foregrounds aforementioned, this polarization leakage is not expected to have a smooth dependence with frequency, which makes the cleaning process more complex \citep{2020/carucci}. 
In this sense, although BINGO is designed to minimize the impact of such effect \citep{2021/wuensche_BINGO-instrument, 2021/abdalla_BINGO-optical_design}, future work will investigate its impact on the foreground cleaning efficiency, as well as on the BAO recovery.

\begin{figure}
  \centering
  \includegraphics[width=\columnwidth]{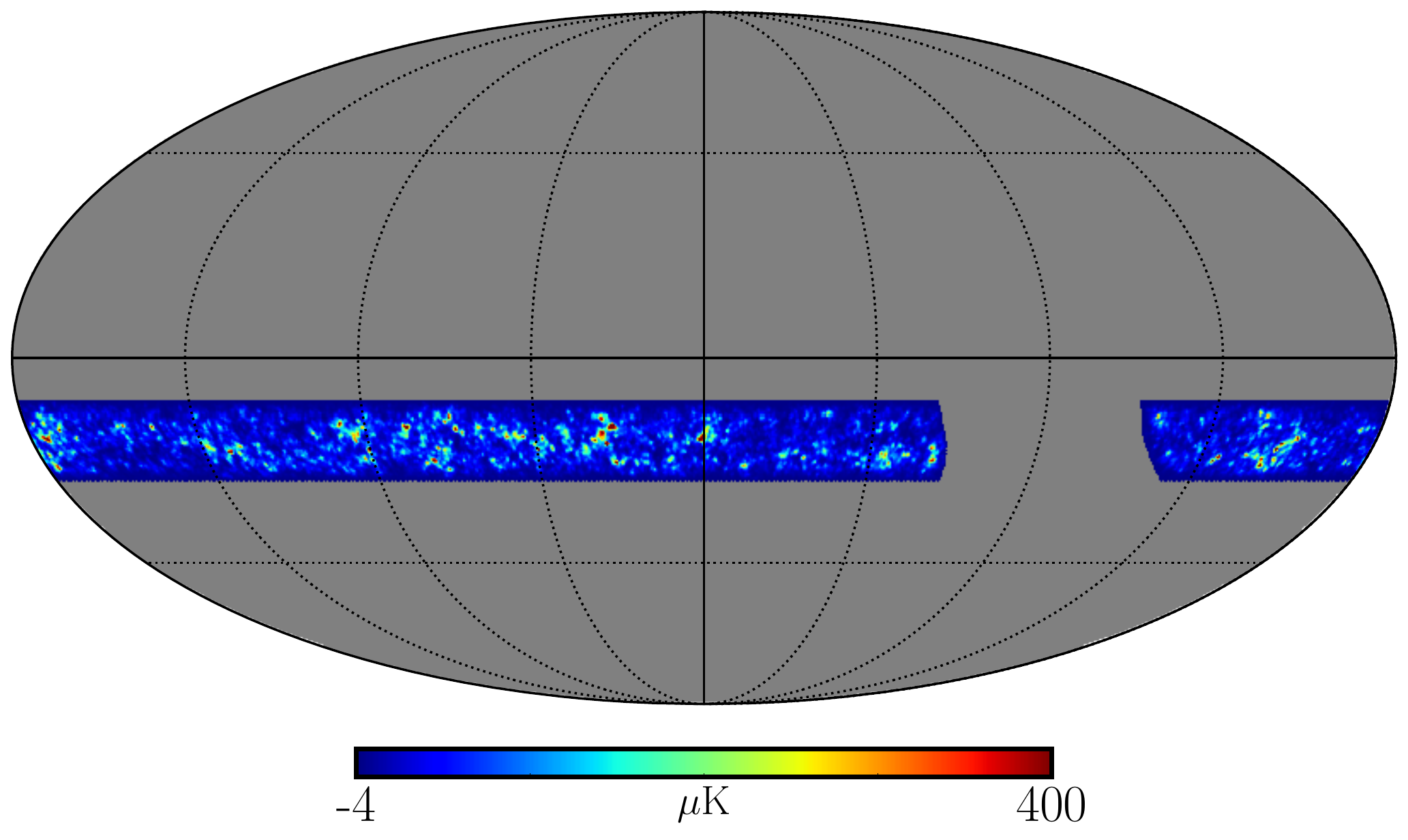}
      \caption{Illustrative example of a {\tt FLASK} simulation, at the smallest redshift bin ($0.127 < z < 0.138$), convolved with a Gaussian beam of $\theta_{\mbox{\sc fwhm}} = 40$ arcmin, after selecting the sky region (apodized mask) considered in all the analyses. The color indicates the brightness temperature in $\mu$K. The Mollweide projection corresponds to celestial coordinates. 
              }
         \label{fig:obs-region}
  \end{figure}

Since all simulations are generated as full sky maps, we select the BINGO region using an appropriate mask, which accounts not only for the expected sky coverage, but also cuts out a region with strong Galactic foreground signal. 
To avoid the impact of sharp edges in the masked region, the mask is apodized with a cosine square transition of 5 degrees using the {\tt NaMaster} code.\footnote{Details about this method can be found in \cite{2019/alonso-namaster}. See also Section \ref{sec:Cls-estimate}.} 
A detailed description of the procedure to create this mask is provided in \cite{2022/mericia_BINGO-component_separation_II}.
Fig. \ref{fig:obs-region} shows an illustrative example of a 21\,cm IM simulation, in the lowest redshift bin $0.127 < z < 0.138$, in the BINGO region. 

We also account for the angular smoothing effect introduced by the BINGO beam, with $\theta_{\mbox{\sc fwhm}} = 40$ arcmin, which we approximate by a Gaussian of same width for all the frequency bins. 
Each simulated sky map is convolved with such a Gaussian beam before thermal noise is added. 
In addition, we investigate the impact of  $\theta_{\mbox{\sc fwhm}}$ scaled with the frequency $\nu$ according to $\theta_{\mbox{\sc fwhm}}(\nu) = \theta_{\mbox{\sc fwhm}}(\nu_0) \times \nu_0/\nu$, where $\nu_0 = 1120$\,MHz and $\theta_{\mbox{\sc fwhm}}(\nu_0) = 40$ arcmin \citep{2015/bigot-sazy}.
Note that, although in both cases we approximate the BINGO beam by a Gaussian, there will be contribution from structures such as the sidelobes, discussed in the companion papers \cite{2021/wuensche_BINGO-instrument} and \cite{2021/abdalla_BINGO-optical_design}, as well as the possible coupling of the horns, still to be investigated.

\subsection{Foreground cleaning process} \label{sec:gnilc}

One of the main challenges of 21\,cm IM observations is the efficient subtraction of the foreground contamination, whose amplitude can be up to $10^4$ larger than the \hi\ signal. 
Indeed, even employing an efficient cleaning technique, it is expected a residual contamination to remain in the observations, so it is important to evaluate their impact on the cosmological analyses. 
In this work, we use the {\tt GNILC} method \citep{2011/remazeilles}, a non-parametric component separation technique, to perform the foreground cleaning process. 
GNILC has been tested in previous work and showed an excellent performance when applied to 21\,cm IM observations \citep{2016/olivari} and, more specifically, to the BINGO simulated data \citep[see papers][]{2021/liccardo_BINGO-sky-simulation, 2021/fornazier_BINGO-component_separation, 2022/mericia_BINGO-component_separation_II}.

The temperature signal, $\mathbf{x}(p)$, represented by a vector of dimension $n_{\rm ch}$ (where $n_{\rm ch}$ is the number of BINGO channels) for each sky pixel $p$, can be modeled as
\begin{equation}
    \mathbf{x}(p) = \mathbf{s}(p) + \mathbf{n}(p) + \mathbf{f}(p) \, ,
\end{equation}
where $\mathbf{s}(p)$ is the 21\,cm signal, $\mathbf{n}(p)$ is the instrumental noise, and $\mathbf{f}(p)$ is the total foreground signal. All contributions to $\mathbf{x}(p)$ have the same dimension $n_{\rm ch}$. 
To distinguish foreground emission from 21\,cm signals, {\tt GNILC} exploits the fact that  foreground emissions are highly correlated between frequencies channels and thus effectively rely on a few independent spectral degrees of freedom, while the 21\,cm signal, probing shells at different redshifts, is only weakly correlated between frequency bands. 
Using a prior (theoretical template) of the 21\,cm and thermal noise power spectra, {\tt GNILC} evaluates, locally in pixel and harmonic space, the effective dimension of the foreground subspace, as identified by a principal component analysis of empirical channel data covariance matrices, computed on a redundant basis (a tight frame) of a type of wavelets on the sphere (called ``needlets''). With this approach {\tt GNILC} does not use any prior information on the foreground components, except the safe assumption of a strong correlation of their emission between frequencies. GNILC then projects the data $\mathbf{x}(p)$ out of the foreground subspace.

The 21\,cm plus noise signal is then reconstructed, for each wavelet scale, applying an ILC filter $\mathbf{W}$ to the data, $\mathbf{\hat{s}} = \mathbf{W x}$, in the subspace orthogonal to that of the foregrounds.
The ILC filter is constructed in such a way to preserve the 21\,cm signal while filtering out the foreground contamination. 
The complete reconstructed 21\,cm plus noise map, at each frequency, is finally obtained by adding the maps corresponding to each wavelet scale.
Detailed discussion on the use of the {\tt GNILC} methodology for 21\,cm foreground cleaning can be found in \cite{2016/olivari} and \cite{2021/fornazier_BINGO-component_separation}.

Finally, it is worth emphasizing that the reconstructed 21\,cm maps contain, along with noise, a residual contribution from the foreground components,
\begin{equation} \label{eq:recHI_plus_residue}
    \mathbf{\hat{s}} \simeq \mathbf{s} + \mathbf{W n} + \mathbf{W f} \, . 
\end{equation}
where $\mathbf{W n}$ and $\mathbf{W f}$ are, respectively, the residual noise and foreground contribution to the reconstructed 21\,cm signal. 
Then, in the case of simulations for which we know the exact input foreground contamination, as is the case here, one can estimate its residual contribution over the recovered cosmological signal. 
Here, we use this facility to include a realistic foreground residual contribution to the simulations to evaluate its impact on the BAO fitting process, avoiding the need for applying {\tt GNILC} to each one of them (see Section \ref{sec:results-flask-foreg}).

\section{Methodology} \label{sec:methodology}

This section summarizes the methodology employed to extract the BAO information from BINGO simulated \hi\ maps.
We describe: 1) how the APS and ACF are measured from the simulated maps; 2) how the covariance matrices are calculated; and 3) the fitting method used to estimate the $\alpha$ parameter from each of the simulations.

\subsection{Clustering measurements}

\subsubsection{Angular power spectrum} \label{sec:Cls-estimate}

The partial-sky coverage is a common situation for CMB and galaxy surveys, whether due to the observation strategy or cuts of regions with strong foreground contamination. 
The APS computation in these cases is not as straightforward as it would be in full-sky analyses, where $C_\ell$'s are computed by expanding the radiation field in spherical harmonics and averaging the $a_{\ell m}$ coefficients: 
\begin{equation}
    C_\ell = \frac{1}{2\ell + 1} \sum_{m=-\ell}^{m=+\ell} |a_{\ell m}|^2 \, .
\end{equation}
The effect of a mask when calculating the $a_{\ell m}$ coefficients over partial sky is to mix (or couple) different multipoles so that the estimated APS, $\hat{C}_\ell$, can be written in terms of the true one, $C_\ell$, through 
\begin{equation}
    \hat{C}_\ell = \sum_{\ell'} \mathcal{M}_{\ell \ell'} C_{\ell'} \, ,
\end{equation}
where $\mathcal{M}_{\ell \ell'}$ is the multipole mixing matrix, which depends on the mask geometry \citep{2002/hivon}.  

Here we estimate the APS using the so-called pseudo-$C_\ell$ formalism, as implemented in the {\tt NaMaster} code\footnote{\url{https://namaster.readthedocs.io}} \citep{2019/alonso-namaster}. 
The mixing matrix is estimated analytically, according to the shape of the mask, and is used to obtain an unbiased estimate of $C_\ell$ \cite[for examples of application of this formalism and/or this code to different cases see, e.g.,][]{2005/brown, 2016/moura-santos, 2019/loureiro, 2020/nicola}. 

However, the loss of information due to the mask usually makes the mixing matrix ill-conditioned and, for this reason, singular, with no direct inversion. 
To deal with that, {\tt NaMaster} allows the usage of bandbowers, calculating a binned and invertible version of $\mathcal{M}_{\ell \ell'}$. 
Here, we consider a linear binning, calculating $C^i_\ell$ at bins of width $\Delta\ell$, centered at $\ell$, a procedure that also helps making the corresponding covariance matrix more diagonal. 
Different $\Delta\ell$ widths are tested so that we can find the more appropriate one to be used in the more realistic analyses, composing the fiducial configuration, defined when running the several robustness tests.

\subsubsection{Angular correlation function}

We measure the ACF from each of the simulations using the following equation \citep{2013/de-simoni}:
\begin{equation} \label{eq:w_theta}
    \omega^i(\theta) = \frac{\sum_{p p'} \delta T_p \delta T_{p'} w t_p w t_{p'}}{\sum_{p p'} w t_p w t_{p'}} \, ,
\end{equation}
with $\delta T_p = T_b - \langle T_b \rangle$, where $T_b$ is the \hi\ brightness temperature at each pixel $p$, $\theta$ is the angular separation between pixels $p$ and $p'$. 
The parameter $w t_p$ is associated to the observed sky region (mask information; Fig. \ref{fig:obs-region}), receiving value 1 in case pixel $i$ contains valid data and 0 in case its area has not been observed or if it is located in the Galactic region and is removed from the analyses.
The $\omega(\theta)$, as given by Eq. (\ref{eq:w_theta}), were calculated at equally spaced $\theta$ values, with bin width $\Delta\theta$, using the public code {\tt TreeCorr}\footnote{\url{https://rmjarvis.github.io/TreeCorr}} \citep{2004/jarvis}. 
As well as for the APS analyses, the best $\Delta\theta$ width for our fiducial configuration will be defined after testing few different values.

\subsection{Covariance matrix} \label{sec:cov-mat}

We use $N=1500$ {\tt FLASK} and $N=500$ N-body mock realizations to estimate the covariance matrices as
\begin{equation}
    \mathtt{C}^{ij}_{km} = \frac{1}{N - 1} \sum_{n=1}^N (X^{i,n}_k - \bar{X}^i_k) \, (X^{j,n}_m - \bar{X}^j_m) \, ,
\end{equation}
where $k$ and $m$ indices run over the different $\ell$ or $\theta$ bins, and $i$ and $j$ indicate the z-bin; $X^i_k$ is the clustering measurement, $C^i_\ell$ or $\omega^i(\theta)$, and $\bar{X}^i_k$ is the respective average over the simulations. 

It is well known that the statistical noise in the covariance matrix estimated from simulations introduces a bias on its inverse, $[\mathtt{C}^{ij}_{km}]^{-1}$, the precision matrix. 
The bias is mainly dependent of the number $N$ of simulations used to construct $\mathtt{C}^{ij}_{km}$ and the number of entries $N_p$ of the data vectors $X$. 
However, as proposed by \cite{2007/hartlap} and used, for example, in the analyses of \cite{2019/camacho, 2018/ata-boss-collab} and \cite{2014/anderson-sdss}, an unbiased version of the precision matrix can be obtained by rescaling it as
\begin{equation}
    [\mathtt{C}^{ij}_{km}]^{-1} \rightarrow \frac{N - N_p - 2}{N - 1} [\mathtt{C}^{ij}_{km}]^{-1} \, .
\end{equation}
The rescaled version of the precision matrix is employed in all our analyses.

\subsection{Parameter inference} \label{sec:param-estimation}

The fit of the template models to the estimated clustering measurement goes through two (recursive) steps: one of them is to fit the nuisance parameters, $B$ and $A_q$, appearing linearly in both template models, and the other is to use the maximum likelihood estimator (MLE) to evaluate  $\alpha$, as described in Eqs. \ref{eq:APS-temp} and \ref{eq:ACF-temp}. 
Following \cite{2018/chan} and \cite{2019/camacho}, we perform a least-square fit to find the best-fit $A_q$ values, and, fixing them, we repeat the procedure to obtain the best-fit $B$ value, in both cases by minimizing the $\chi^2$ defined as
\begin{equation}
    \chi^2(\lambda) = \sum_{k,m} [ X^i_k - x^i_k(\lambda) ] \, [\mathtt{C}^{ij}_{km}]^{-1} \, [ X^j_m - x^j_m(\lambda) ] \, ,
\end{equation}
where $\lambda = \{\alpha, B, A_q\}$ and $x^i_k$ is the template model for the $i$th z-bin and the $k$th $\ell$/$\theta$ bin. 
Having done that, we maximize the likelihood function, 
\begin{equation}
    L(\alpha) \propto \exp(-\chi^2 / 2) \, , 
\end{equation}
now dependent on only one parameter, since $\chi(\lambda)$ is reduced to $\chi(\alpha)$, allowing to estimate the best-fit $\alpha$ value \cite[see also,][]{2014/anderson-sdss}. 
In practice, this value is obtained through a grid search, so that the least-square fit of the nuisance parameters, $A_q$ and $B$, is repeated for each $\alpha$ value evaluated.

Taking advantage of a tomographic approach, we are able to combine redshift bins so that the significance of the BAO detection can increase. 
We fit one $\alpha$ parameter through a joint analysis applying the MLE over a set of $N_z$ consecutive z-bins. 
In particular, we fit the nuisance parameters, $B$ and $A_q$, to each z-bin individually, and, fixing their values, we find one best-fit $\alpha$ for each set of $N_z$ z-bins. 

The $1\sigma$ error for the $\alpha$ estimates, $\sigma_\alpha$, is defined as the deviation from the maximum likelihood point, at $\chi_{min}^2$, by $\Delta \chi^2 = 1$. 
It is worth noting that this $\Delta \chi^2 = 1$ rule is valid in the case of a Gaussian likelihood, $L(p)$, for only one parameter, in our case $p=\alpha$ \citep[see][and references therein, for details]{2018/chan}. 
In principle, this can be employed in our analyses, as commonly seen in literature \citep[e.g.,][]{2018/ata-boss-collab, 2018/carter-6df, 2019/abbott-des, 2021/abbott-des}, but one has to be careful with deviations from Gaussian distribution, which can lead to $\sigma_\alpha$ values that underestimate the error bars.
In the next section we discuss and evaluate the validity of this rule for each clustering statistic used here.

\section{BAO fitting results} \label{sec:results}

In this section we present our findings from analyzing {\tt FLASK} and N-body mock realizations. 
We start with an ideal case, analyzing pure 21\,cm simulations, only accounting for the BINGO sky coverage and fixed 40 arcmin beam (Section \ref{sec:results-21cm-only}). 
The contribution of thermal noise and foreground residuals are then considered one at a time, so that we can investigate the impact of each one separately, as well as the impact of a redshift dependent beam size (Sections \ref{sec:results-flask-noise} and \ref{sec:results-flask-foreg}). 
We also present the results from an extensive (but not exhaustive) list of robustness tests evaluating several aspects of the methodology employed here to extract the BAO signal from BINGO-like simulations, using {\tt FLASK} mocks as the cosmological signal (Section \ref{sec:robustness-tests}). 
Such tests allow finding the most appropriate configuration (hereafter called fiducial configuration), optimizing the analysis. 
In particular, the fiducial configuration is defined by: 
\begin{enumerate}
    \item[(a)] angular and multipole binning: $\Delta \theta = 0.5^\circ$ and $\Delta \ell = 10$, 
    \item[(b)] $A_q$ parameters for $C_\ell$ and $\omega(\theta)$ templates: $q = -1, 0, 1, 2$ and $q = 0, 1, 2$, respectively. 
\end{enumerate}

\noindent
The 500 N-body mocks are analyzed using only this fiducial configuration (Section \ref{sec:mock-results}). 

Also important is the choice for the range of scales, $\theta$ and $\ell$ intervals. 
Since BINGO coverage is limited to a declination strip of $\sim 15^\circ$, the $C_\ell$ determination at the smallest multipoles ($\ell \lesssim 15$) is compromised. 
Then, applying the BAO fitting to the (high signal-to-noise ratio) mean $C_\ell$ from the 1500 {\tt FLASK} mocks we chose $\ell_{min} \approx 32$, for which $\alpha$ estimate is the least biased with the smallest $\sigma_\alpha$ error.
This way the choice of multipole range can also account for any artifact introduced by the mask geometry, which would be more evident in the mean clustering measurement.
Details about the fitting procedure over the mean APS and ACF are presented in the next subsections. 

\begin{figure}[h]
\centering
\includegraphics[width=1\columnwidth]{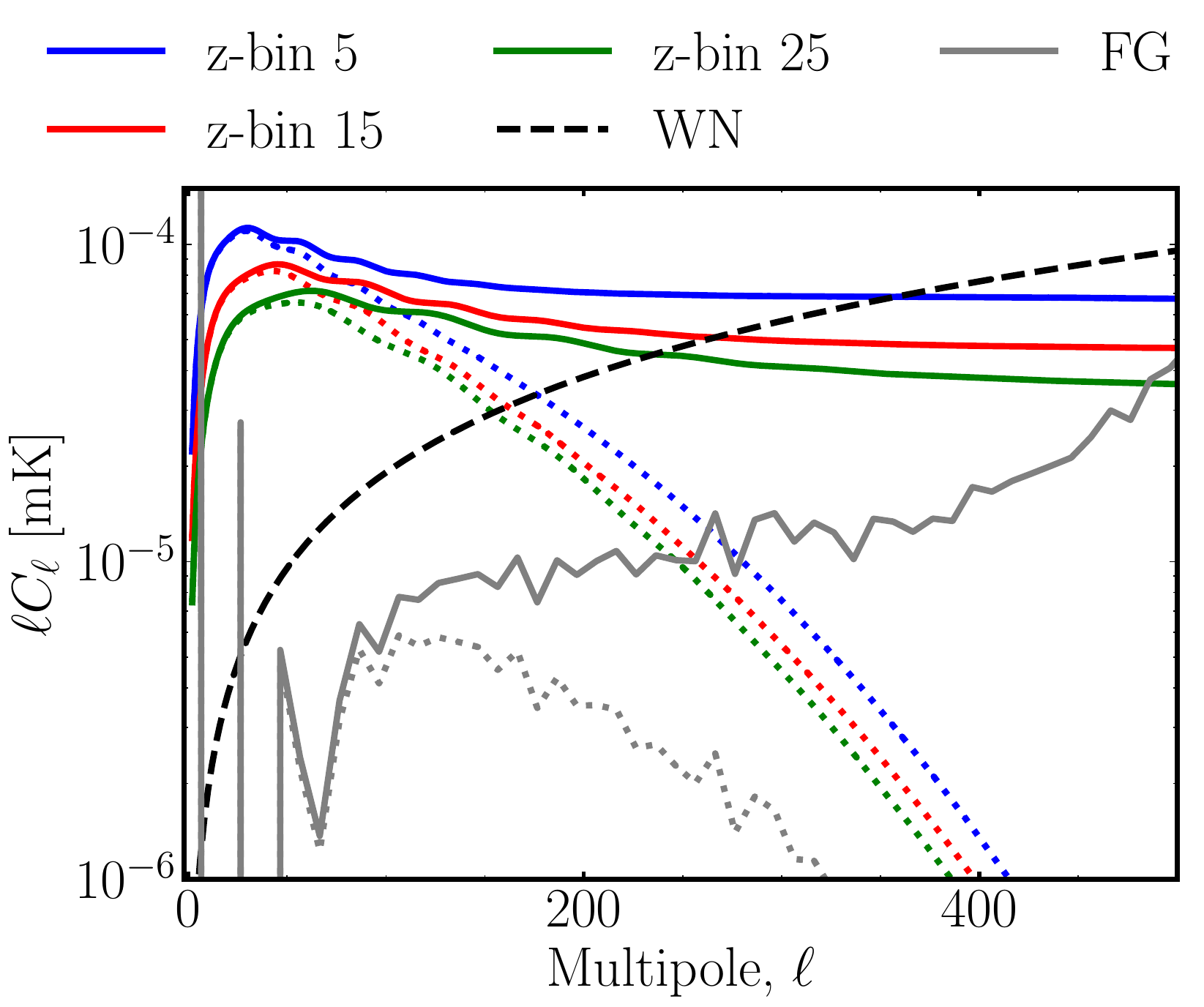}
\caption{Theoretical APS calculated for three of the 30 tomographic bins linearly spaced in frequency, whose average redshifts are $\bar{z}_5 = 0.166$, $\bar{z}_{15} = 0.263$ and $\bar{z}_{25} = 0.377$. Note the BAO wiggles appearing at higher multipoles the higher the redshift. The black dashed line shows the white noise amplitude. The expected contribution from foreground residual at the highest z-bin ($\bar{z}_{30} = 0.442$; the one with the highest amplitude of foreground residual) is represented by the gray line. The dotted lines show the effect introduced by the instrumental beam with $\theta_{\mbox{\sc fwhm}} = 40'$.  }
\label{fig:Cl-Nl}
\end{figure}

For all the z-bins we consider the same minimum multipole, but the maximum one is chosen according to the multipole ranges at which the BAO wiggles are concentrated, which are more spread out over higher multipoles the higher the redshift (Fig. \ref{fig:Cl-Nl}). 
In this sense, after few tests over the mean $C_\ell$, we chose $141 \lesssim \ell_{max} \lesssim 401$ for lower to higher redshifts, so that we use only the range containing the BAO wiggles, avoiding to increase the number of data points with multipoles that do not bring information on the BAO signal (see discussion in Sec. \ref{sec:cov-mat}). 

A similar procedure is used to choose the $\theta$ range for the ACF analyses. 
Applying the BAO fitting pipeline over the mean $\omega(\theta)$ measurements from the {\tt FLASK} mocks we chose, for each z-bin, the range of scales large enough to encompass a bit more than the full width of the BAO peak. 
Our choices for the $\theta$ ranges goes from $10.5^\circ \lessapprox \theta \lessapprox 21.0^\circ$, for the smallest z-bin, to $2.5^\circ \lessapprox \theta \lessapprox 8.0^\circ$, for the highest z-bin. 

The $\theta$ and $\ell$ ranges used for each z-bin in all our analyses are presented in Table \ref{tab:theta_ell_ranges}. 
The large number of z-bins demanded a large number of tests to find the more appropriate angular and mutipole ranges for each estimator, reason why we only show our final choice in each case.
We use the \textit{approximate} ($\approx$) symbol to refer to minimum and maximum $\theta$ and $\ell$ values because, although these numbers are exact for the fiducial values of $\Delta \theta$ and $\Delta \ell$, they are slightly different when testing other binning schemes. 
For such cases we choose the range of scales as close as possible to the fiducial configuration, shown in Table \ref{tab:theta_ell_ranges}. 
Note that all tests to choose the best angular and multipole ranges are applied to the BINGO-like simulations, since our goal is optimize our analyses for future BINGO observations. 

\begin{table}[h!]
\linespread{1.3}
\setlength{\tabcolsep}{5pt}
\selectfont
\centering
\caption{Maximum multipole, $\ell_{max}$, and minimum and maximum angular scales, $\theta_{min}$ and $\theta_{max}$, in degrees, considered for the template fitting over APS and ACF clustering estimates from each z-bin. For the APS, the minimum multipole is the same for all the z-bins, $\ell_{min} \approx 32$. The tomographic bins, from lower to higher redshifts, are enumerated from 1 to 30. The central redshift of each z-bin, $\bar{z}$, is also shown.} 
{\small
\begin{tabular}{|c|c|c|c||c|c|c|c|}
\hline
\# & $\bar{z}$ & $\ell_{max}$ & $[\theta_{min}, \theta_{max}]$ & \# & $\bar{z}$ & $\ell_{max}$ & $[\theta_{min}, \theta_{max}]$\\
\hline
1  & 0.131 & 141 & [10.5, 21.0] & 16 & 0.274 & 301 & [5.0, 10.5] \\
2  & 0.140 & 141 & [10.5, 20.0] & 17 & 0.284 & 311 & [4.5, 10.5] \\
3  & 0.148 & 151 & [9.0, 18.5]  & 18 & 0.295 & 321 & [4.5, 10.0] \\
4  & 0.157 & 161 & [8.5, 17.5]  & 19 & 0.306 & 331 & [4.0, 10.0] \\
5  & 0.166 & 181 & [8.0, 17.0]  & 20 & 0.318 & 341 & [4.0, 10.0] \\
6  & 0.175 & 201 & [7.5, 16.0]  & 21 & 0.329 & 361 & [4.0, 9.5]  \\
7  & 0.184 & 211 & [7.0, 15.5]  & 22 & 0.341 & 371 & [3.5, 9.5]  \\
8  & 0.194 & 231 & [7.0, 15.0]  & 23 & 0.353 & 371 & [3.5, 9.5]  \\
9  & 0.203 & 251 & [6.5, 14.0]  & 24 & 0.365 & 381 & [3.5, 9.0]  \\
10 & 0.213 & 251 & [6.5, 13.5]  & 25 & 0.377 & 391 & [3.0, 9.0]  \\
11 & 0.222 & 271 & [6.0, 13.0]  & 26 & 0.390 & 391 & [3.0, 9.0]  \\
12 & 0.232 & 271 & [6.0, 12.5]  & 27 & 0.403 & 401 & [3.0, 8.0]  \\
13 & 0.242 & 271 & [5.5, 11.5]  & 28 & 0.416 & 401 & [3.0, 8.0]  \\
14 & 0.252 & 281 & [5.5, 11.5]  & 29 & 0.429 & 401 & [2.5, 8.0]  \\
15 & 0.263 & 291 & [5.0, 11.0]  & 30 & 0.442 & 401 & [2.5, 8.0]  \\
\hline 
\end{tabular} }
\label{tab:theta_ell_ranges}
\end{table}

\subsection{Fitting clustering measurements from 21\,cm only simulations} \label{sec:results-21cm-only}

Employing the fiducial configuration, the BAO signal is measured by jointly fitting three sets of $N_z = 10$ consecutive $z$-bins, the lower, intermediate and higher $z$-bins (redshift ranges of $\Delta z \approx 0.09, 0.11, 0.12$ widths), estimating only one $\alpha$ parameter for each set. 
The advantage of a tomographic approach, combining different z-bins, is to improve the error bars and statistical significance of the measurements.
Notice that the nuisance parameters, one parameter $B$ and 4 (3) parameters $A_q$, are fitted to the $C_\ell$ ($\omega(\theta)$) estimated for each of the 10 z-bins individually, so that, for the fiducial configuration, we have a total of $10 \times 5 (4) = 50 (40)$ nuisance parameters. 
The $\alpha$ fitting results obtained from analyzing the 21\,cm  only simulations (accounting for BINGO sky coverage and beam) are shown in the first part of Tables \ref{tab:results-21cmWNFG-ACF} and \ref{tab:results-21cmWNFG-APS} for the ACF and APS estimators, respectively. 

We start testing our BAO fitting pipeline over the high signal-to-noise ratio mean clustering measurements, that is, the mean $C_\ell$ and $\omega(\theta)$ from the 1500 {\tt FLASK} mocks. 
Our aim in fitting such high signal-to-noise measurement is not only to help decide about angular and multipole ranges and their binning widths, $\Delta\ell$ and $\Delta\theta$, but also to be able to identify any possible systematic bias on $\alpha$ estimates. 
Hereafter, we use $\alpha^m$ and $\sigma^m_\alpha$ to denote the results of fitting the mean $C_\ell$ or $\omega(\theta)$, and $\alpha$ and $\sigma_\alpha$ the results for individual mocks. 
The fitting over the mean clustering measurement requires the original covariance matrix to be divided by the total number of simulations, but such scaling would not provide a realistic estimate of $\sigma_\alpha$ for one mock realization.
Therefore, since likelihoods can be well approximated by Gaussians for high signal-to-noise ratio measurements of the BAO feature, we follow \cite{2019/abbott-des} and divide the original covariance matrix by 10 to obtain the likelihood and fit the $\alpha^m$ parameter. 
The corresponding $1\sigma$ uncertainty is calculated as described in the previous section and, then, normalized to obtain $\sigma^m_\alpha \rightarrow \sqrt{10}\sigma^m_\alpha$. 
Notice that using a different normalization factor, for example, 20 instead of 10, leads to similar error amplitude. 
The $\alpha^m \pm \sigma^m_\alpha$ results are presented in the last column of Tables \ref{tab:results-21cmWNFG-ACF} and \ref{tab:results-21cmWNFG-APS}. 
Using the fiducial configuration to fit the mean clustering measurements from the 21 cm only simulations we find a prediction of 24\%, 10\%, and 5\% errors for the lower, intermediate and higher z-bins, respectively. 
Note that the same procedure is employed to all tests fitting the mean clustering measurements, namely, varying the fiducial configuration or/and including contaminant signals, whose results are shown in the last column of Tables \ref{tab:results-21cmWNFG-ACF} to \ref{tab:results-21cmWNFG-APS-mocks} and discussed in the next subsections. 
In general, $\alpha^m$ estimates are consistent with unity, although presenting small bias, which, in all the cases, is well below the statistical uncertainty.

\begin{figure*}[h]
\includegraphics[width=0.75\textwidth]{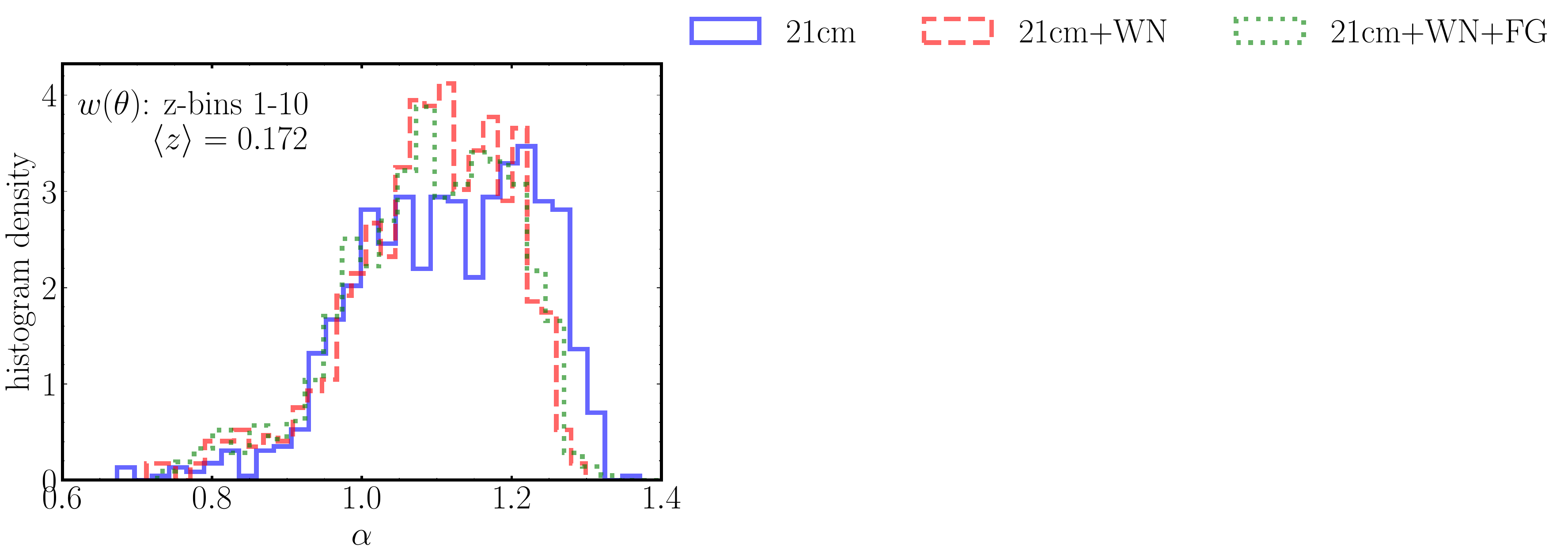}\hspace*{-7.85cm}\includegraphics[width=0.34\textwidth]{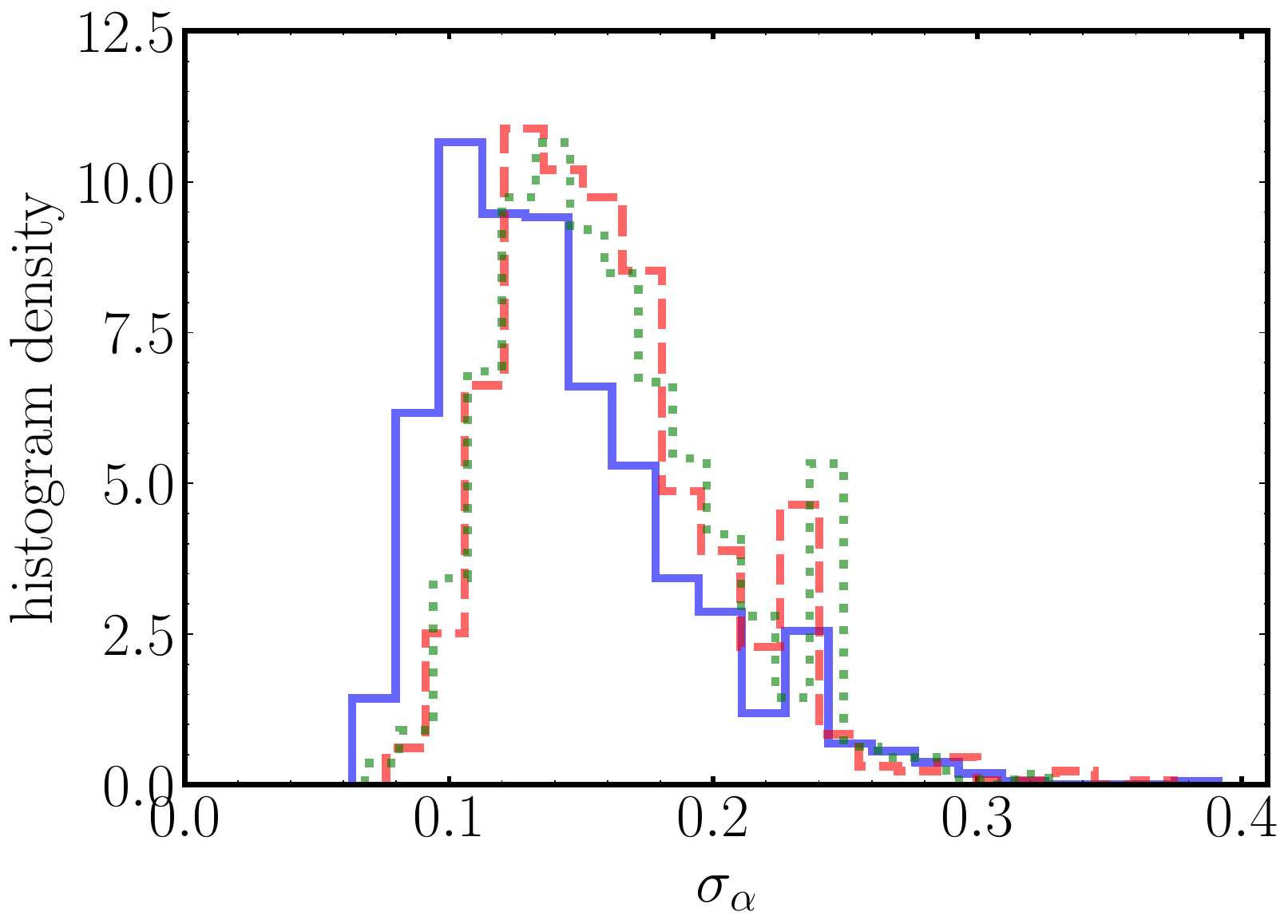}\includegraphics[width=0.33\textwidth]{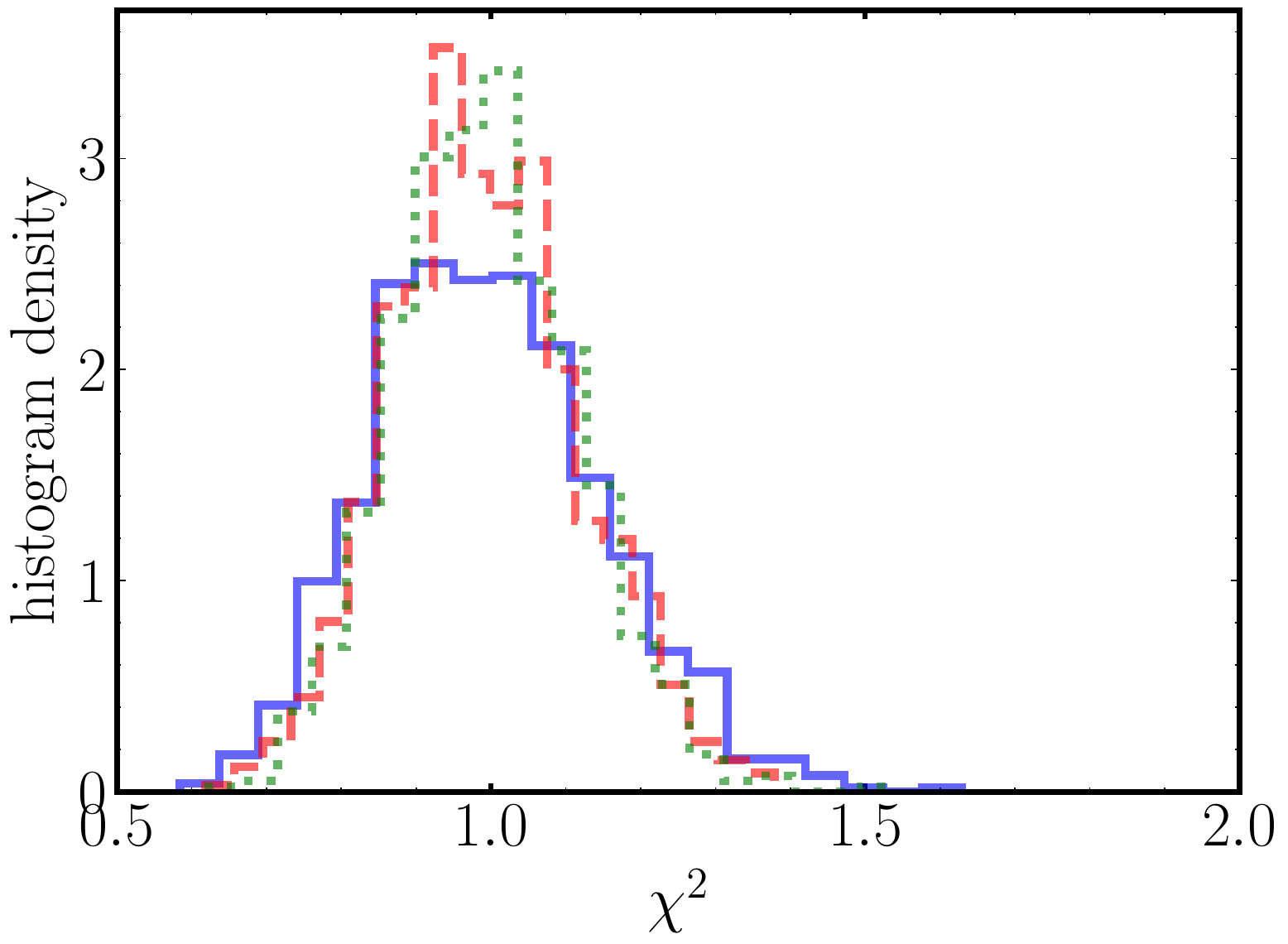}\\
\vspace{-0.85cm}\\
\includegraphics[width=0.33\textwidth]{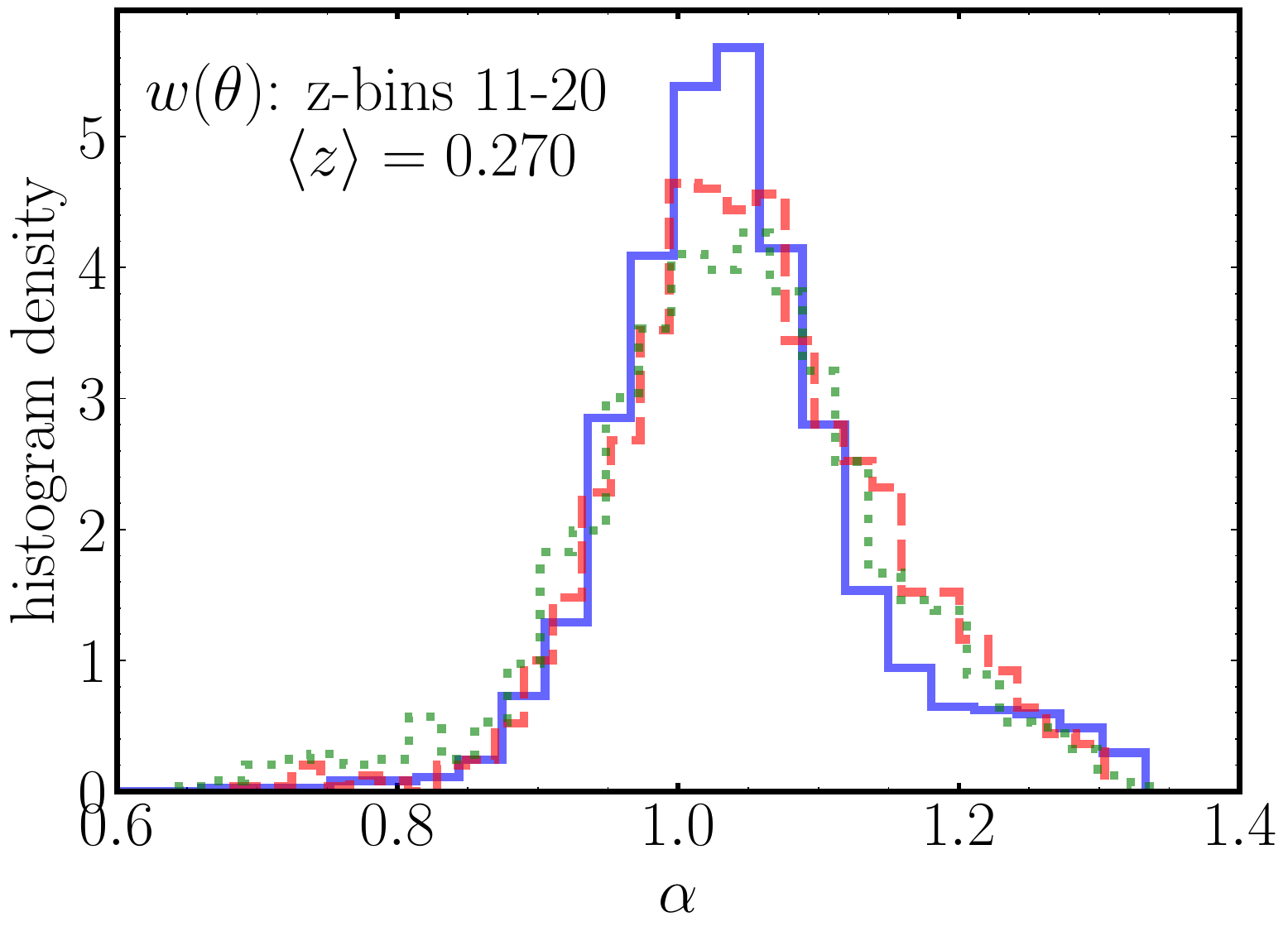}\includegraphics[width=0.33\textwidth]{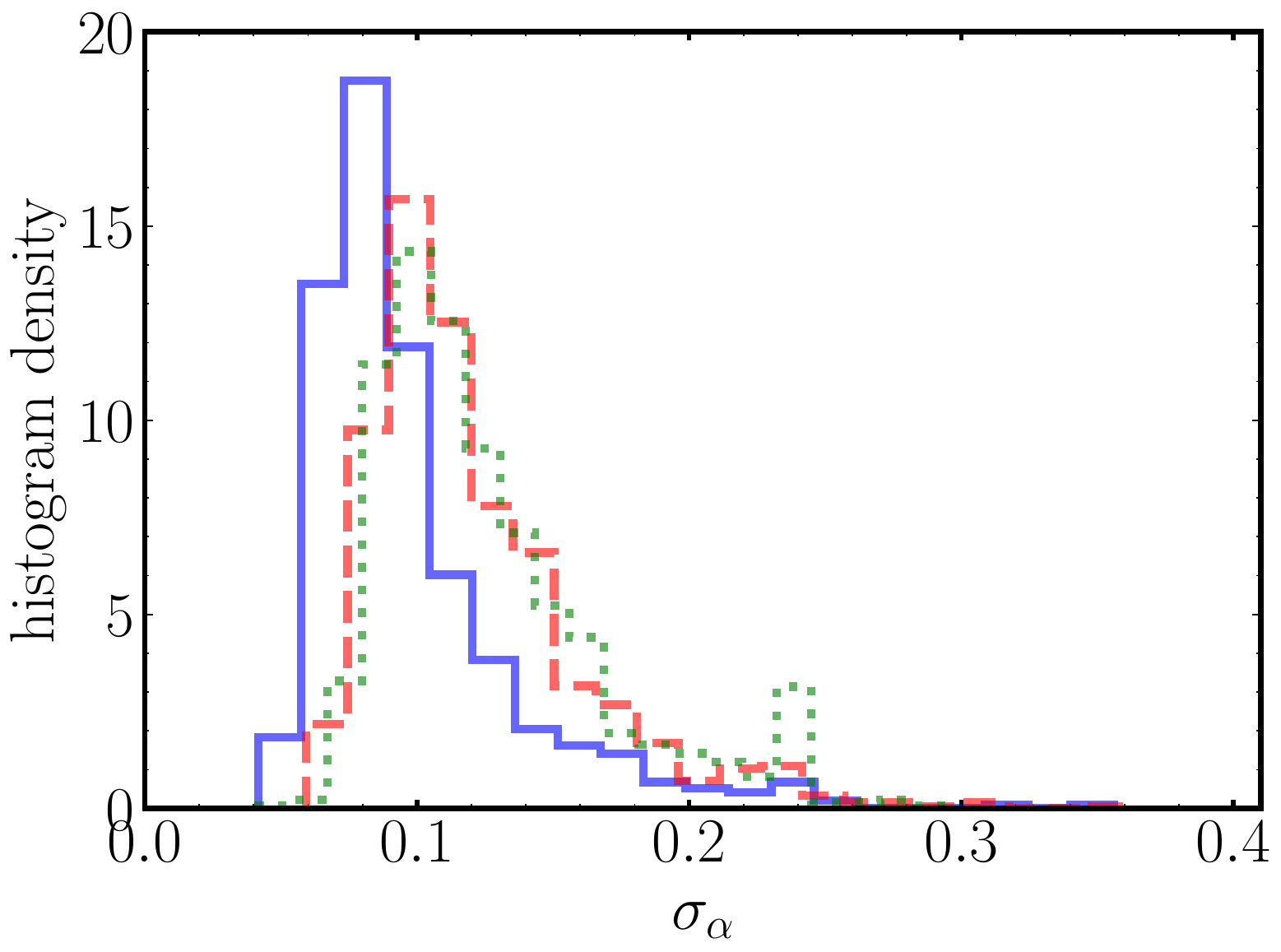}\includegraphics[width=0.34\textwidth]{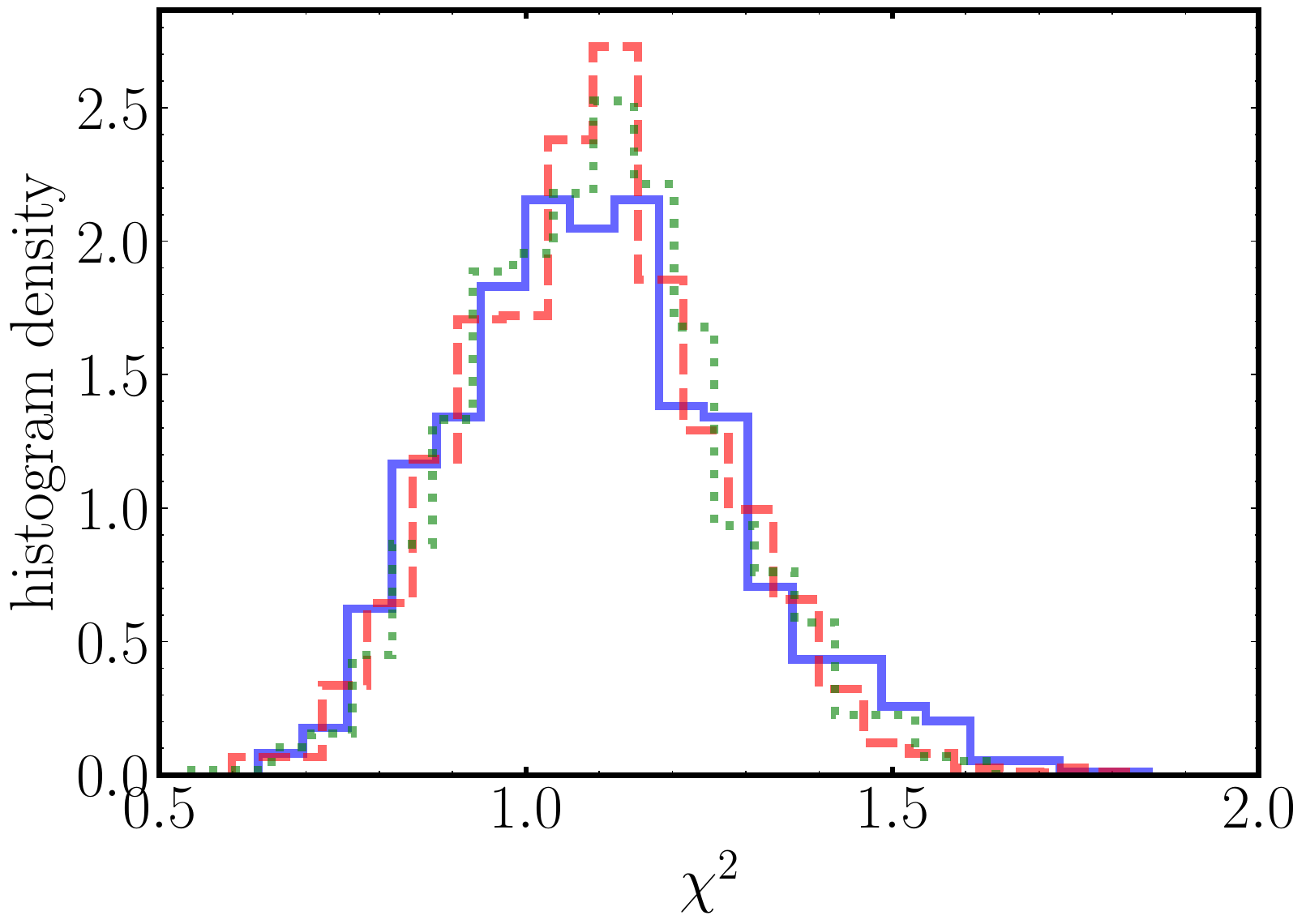}\\
\vspace{-0.85cm}\\
\includegraphics[width=0.33\textwidth]{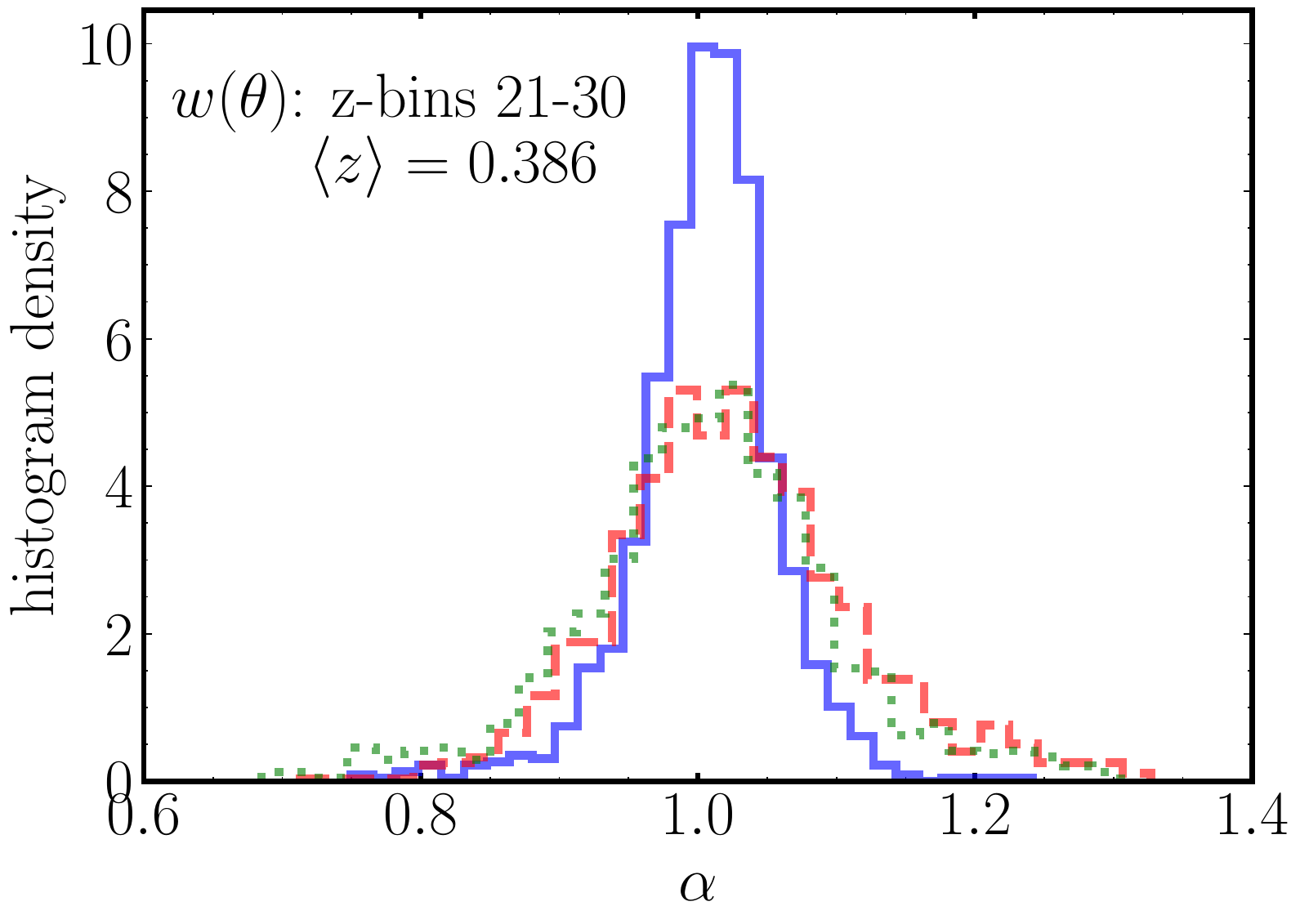}\includegraphics[width=0.33\textwidth]{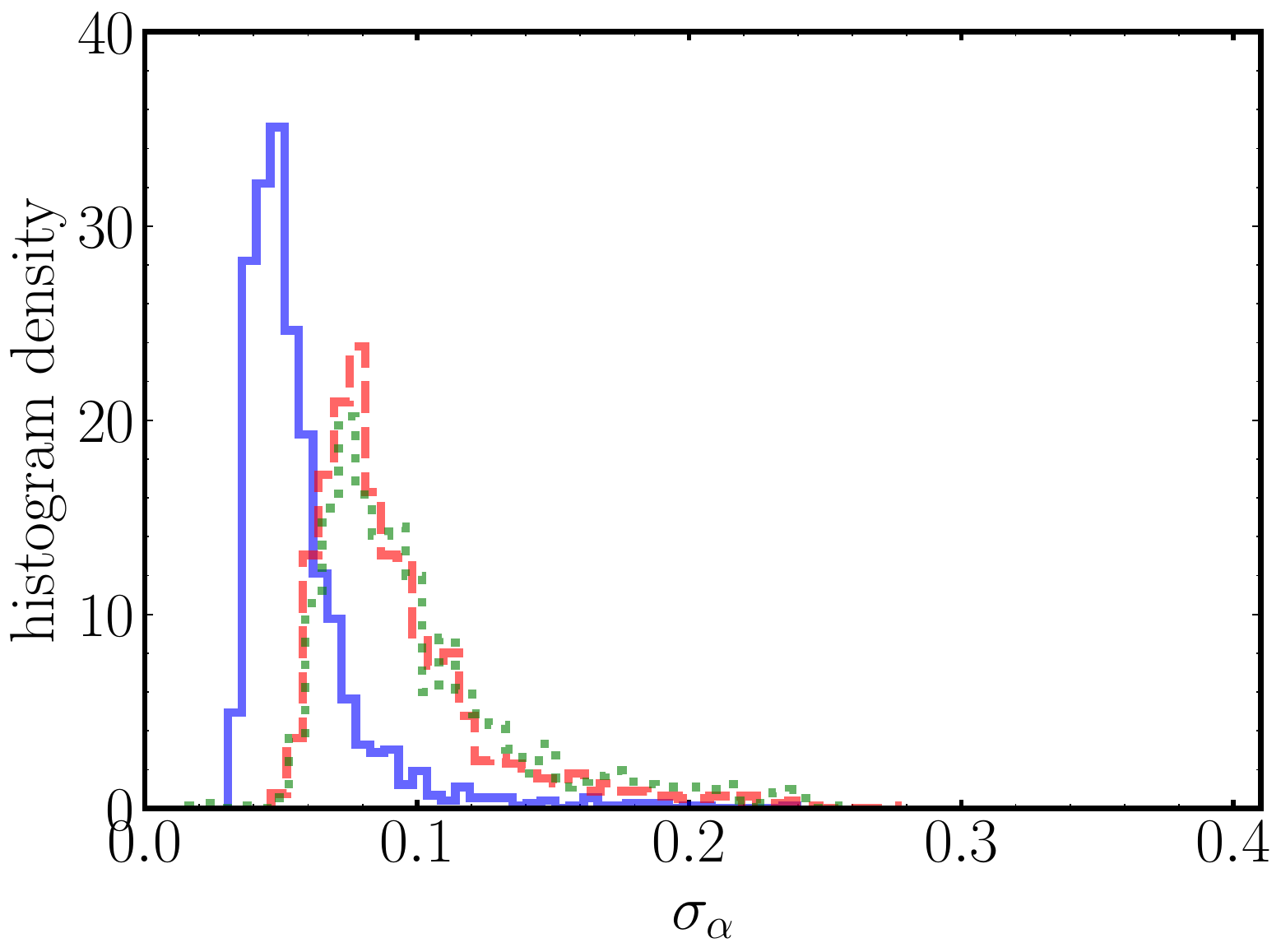}\includegraphics[width=0.34\textwidth]{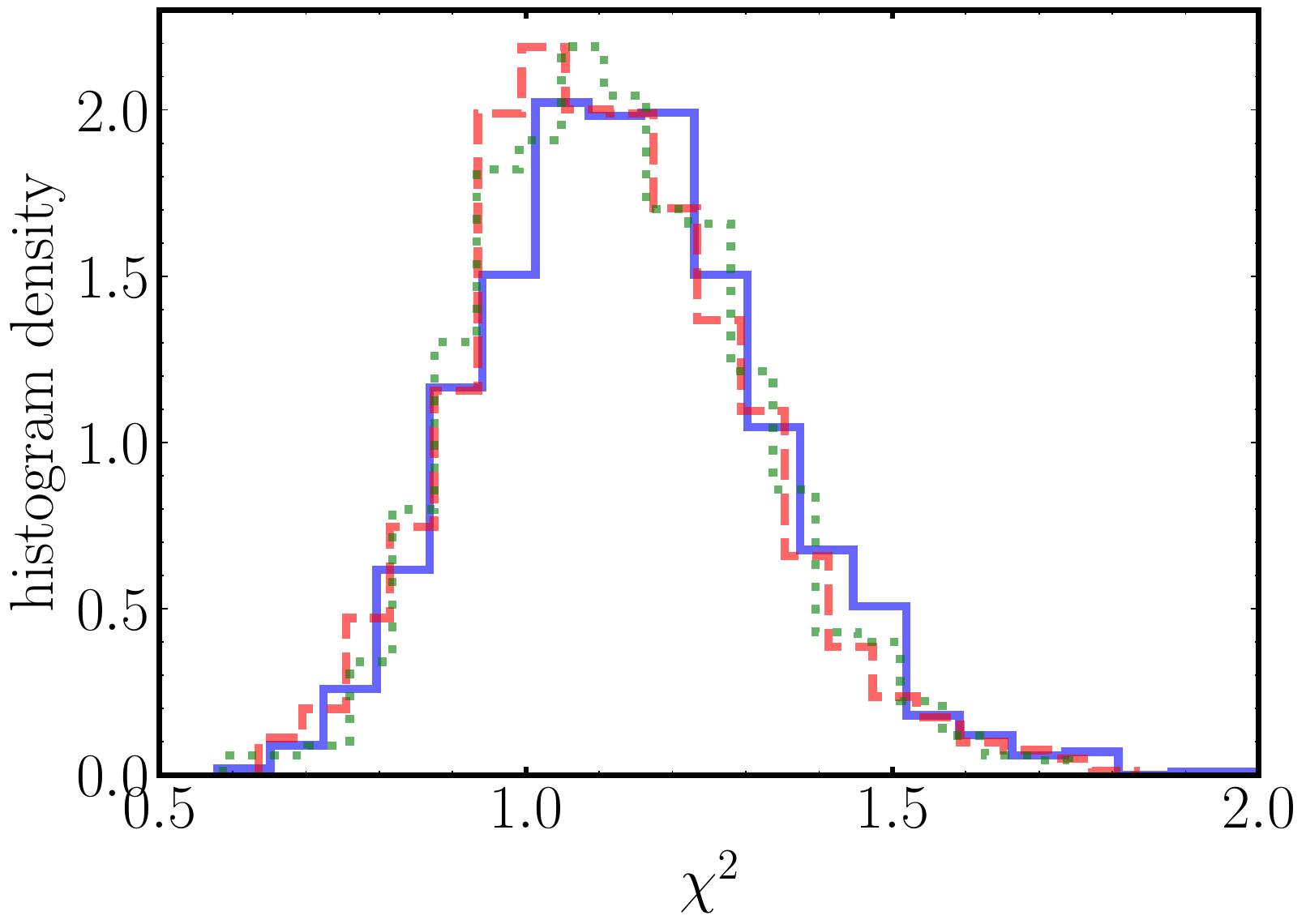}
\caption{Histogram distribution of the $\alpha$ parameter, $\sigma_\alpha$ error, and $\chi^2$ obtained applying the BAO fitting pipeline to each of the {\tt FLASK} simulations for the ACF estimator using the fiducial configuration. 
Rows from top to bottom show results for each redshift range, the lower (1-10 z-bins), the intermediate (11-20 z-bins), and the higher redshifts (21-30 z-bins). The average redshift for each range is presented in the first panel of each row. 
In all the plots, the solid blue, dashed red and dotted green lines show the distribution of the parameters resulting from analyzing, respectively, the 21\,cm only simulations (21\,cm), the noisy 21\,cm simulations (21\,cm + WN), and the BINGO-like simulations (adding also the foreground residual; 21\,cm + WN + FG). All these results correspond to the $N_s$ fraction of the mocks. See Table \ref{tab:results-21cmWNFG-ACF}.}
\label{fig:baoResults-ACF}
\end{figure*}

\begin{figure*}[h]
\includegraphics[width=0.75\textwidth]{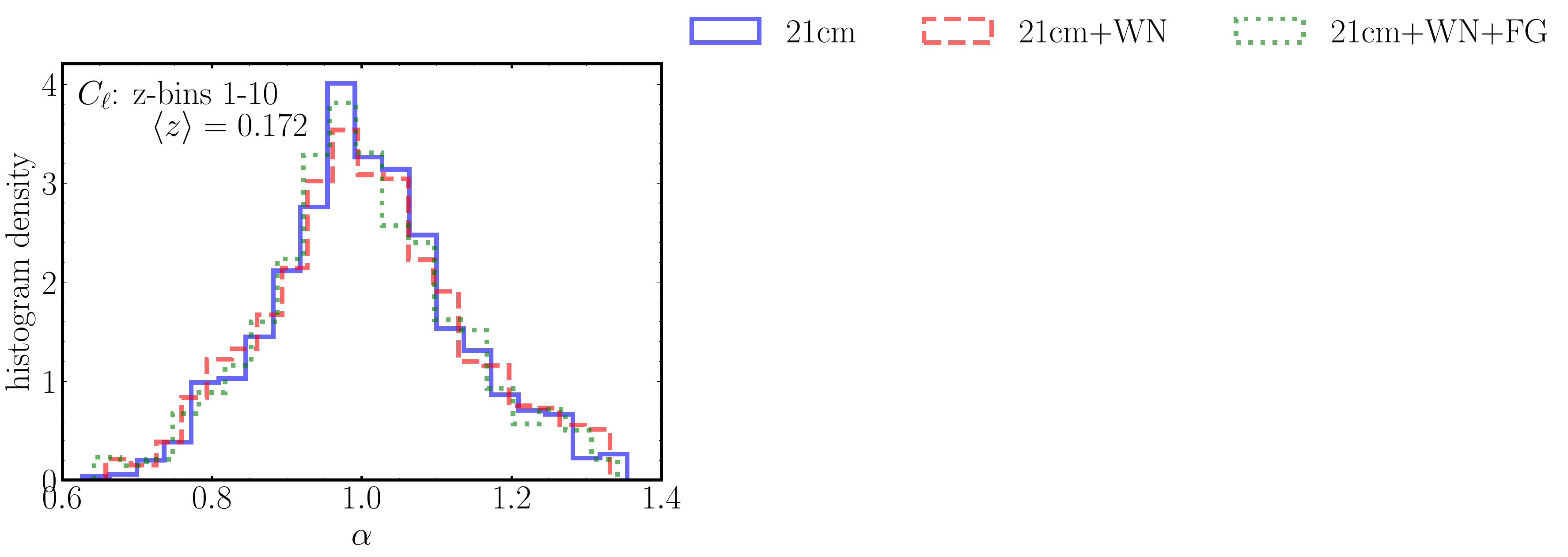}\hspace*{-7.8cm}\includegraphics[width=0.335\textwidth]{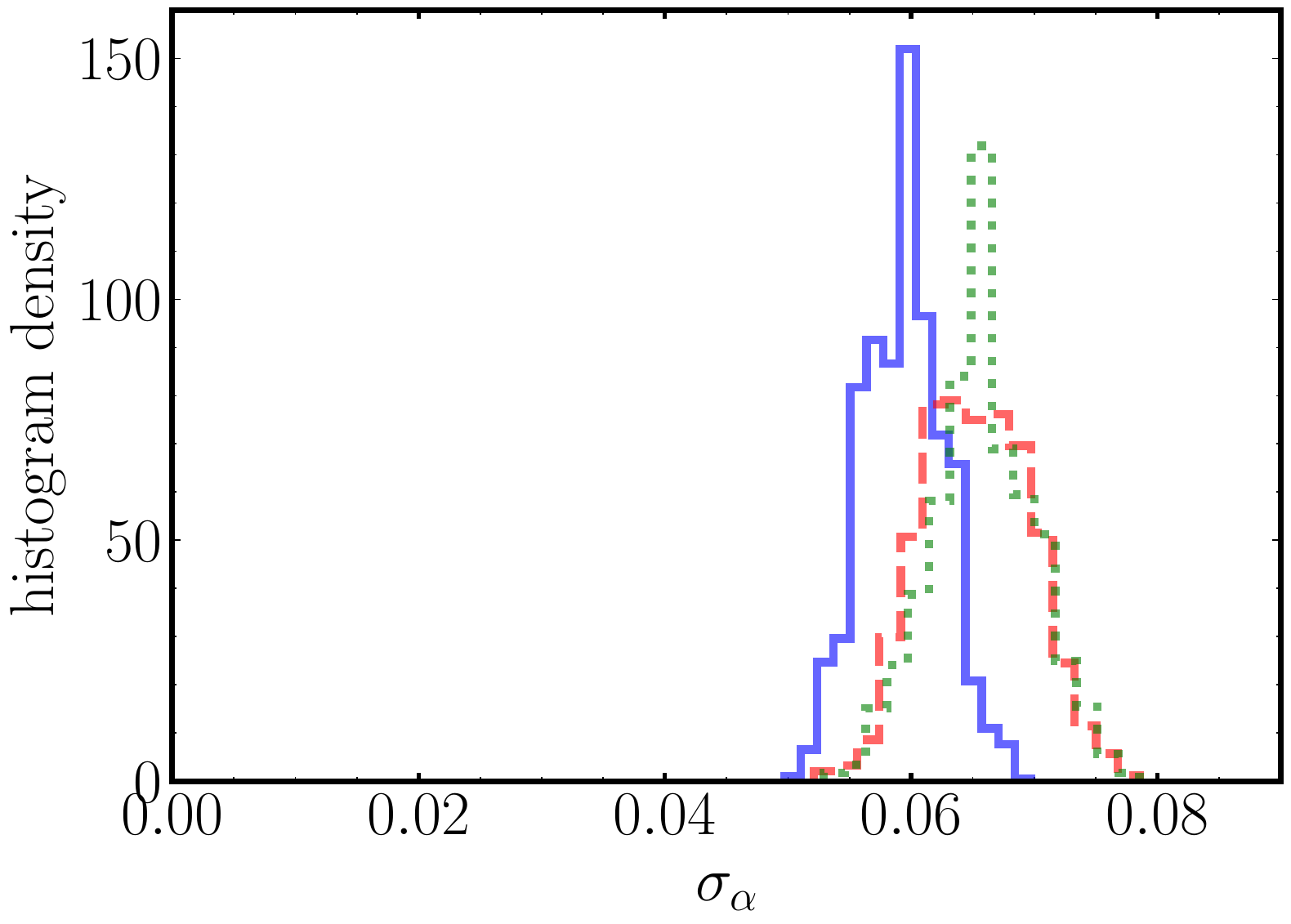}\includegraphics[width=0.34\textwidth]{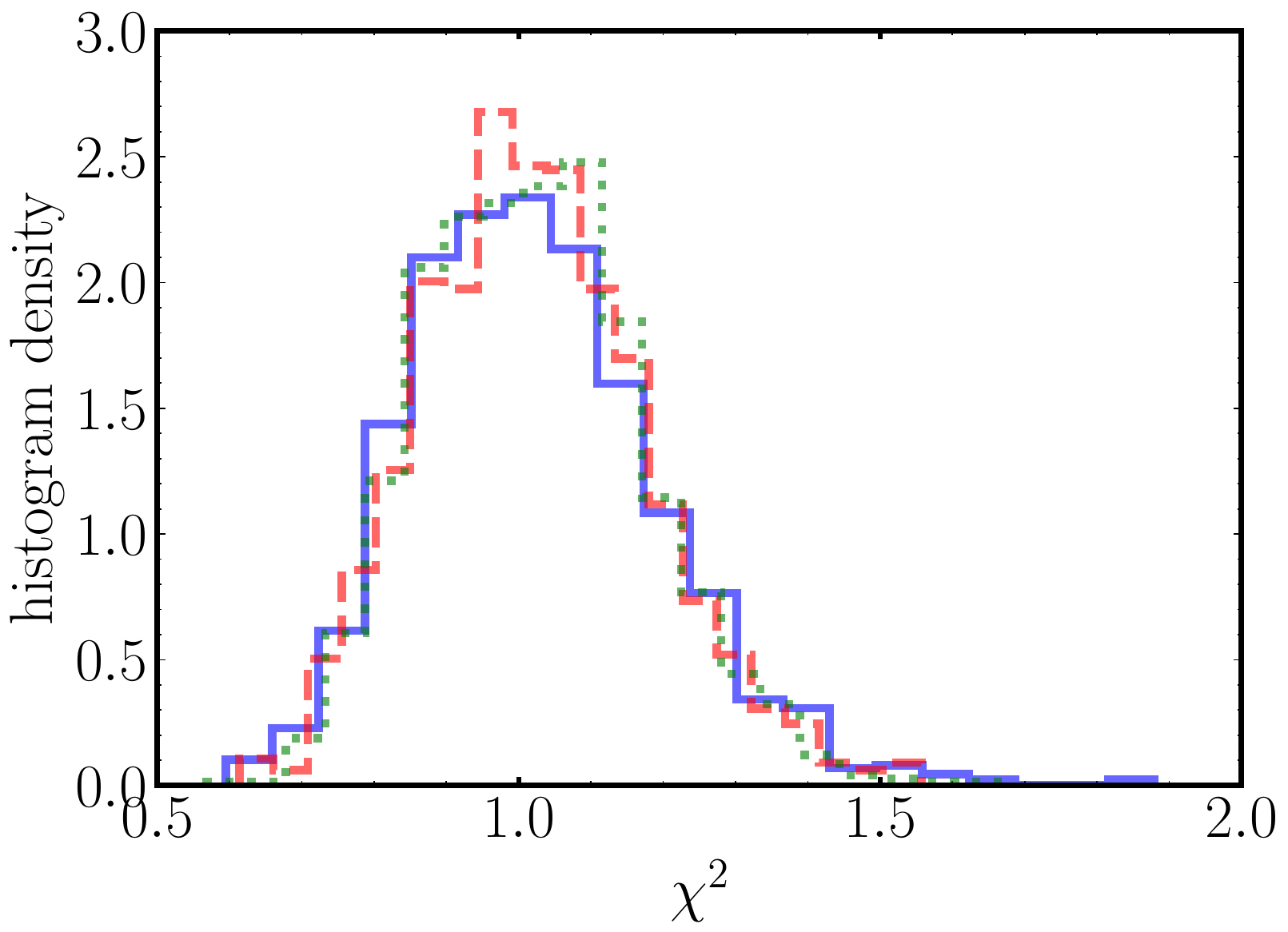}\\
\vspace{-0.8cm}\\
\includegraphics[width=0.33\textwidth]{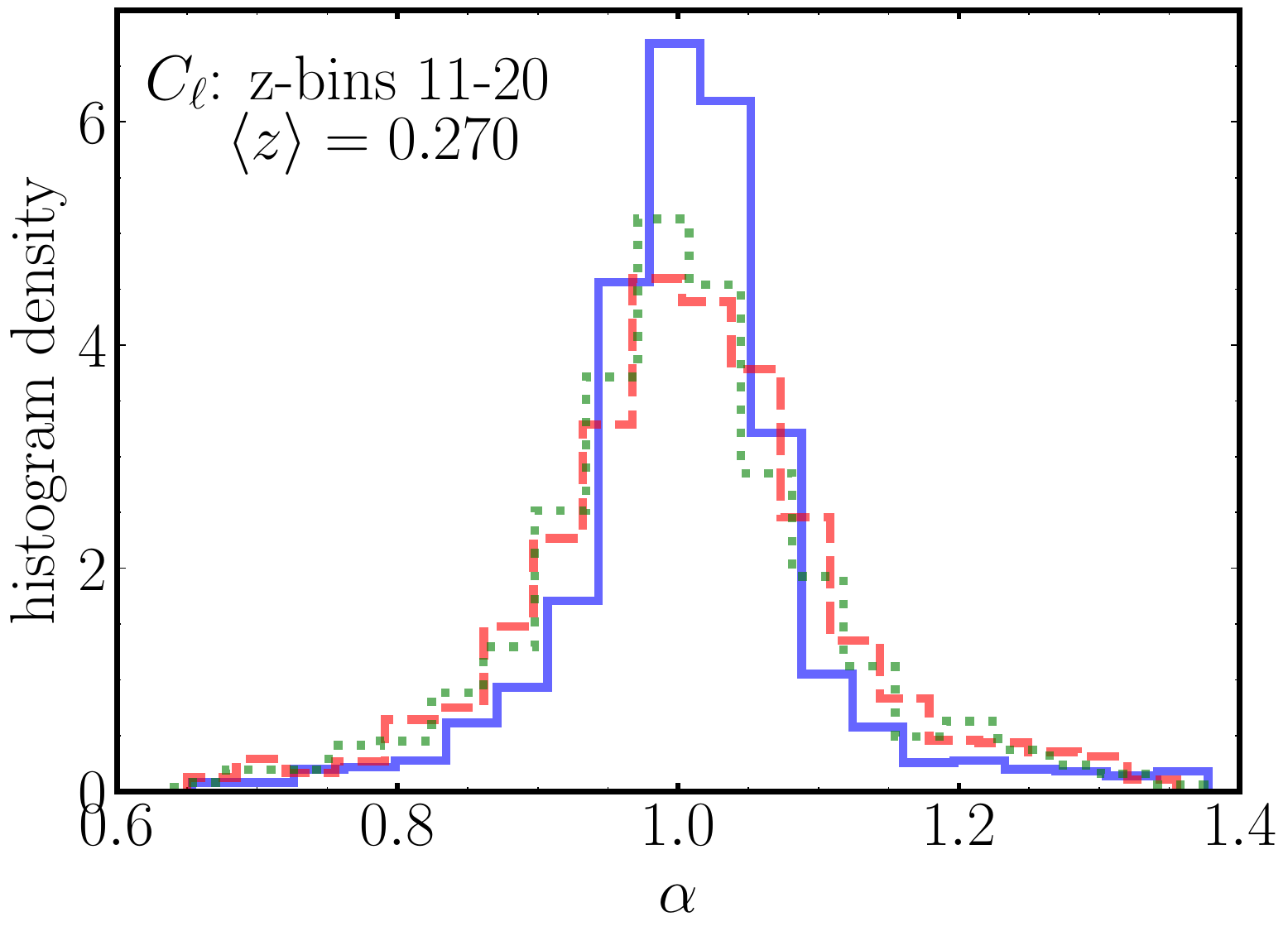}\includegraphics[width=0.33\textwidth]{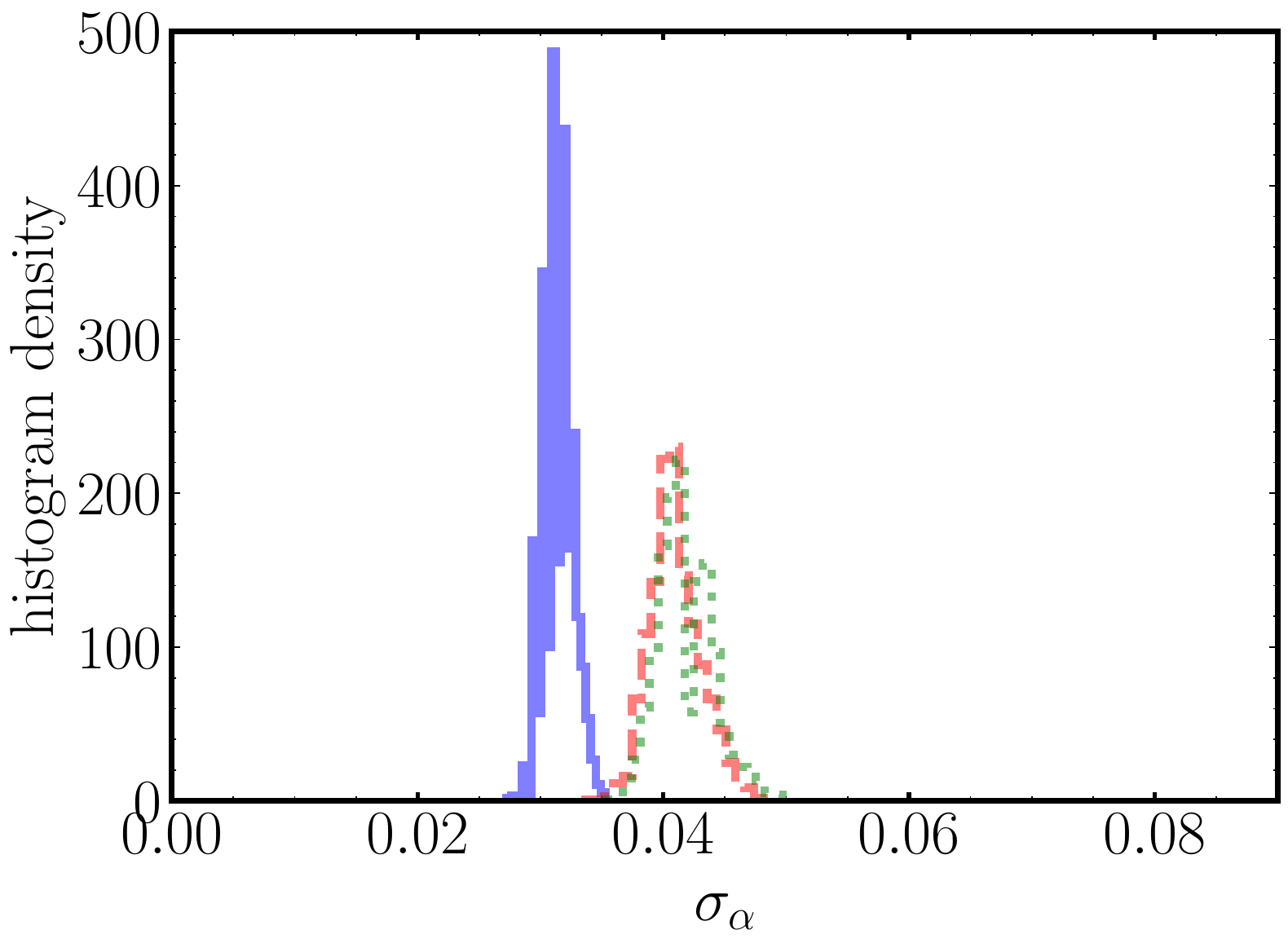}\hspace*{0.25cm}\includegraphics[width=0.33\textwidth]{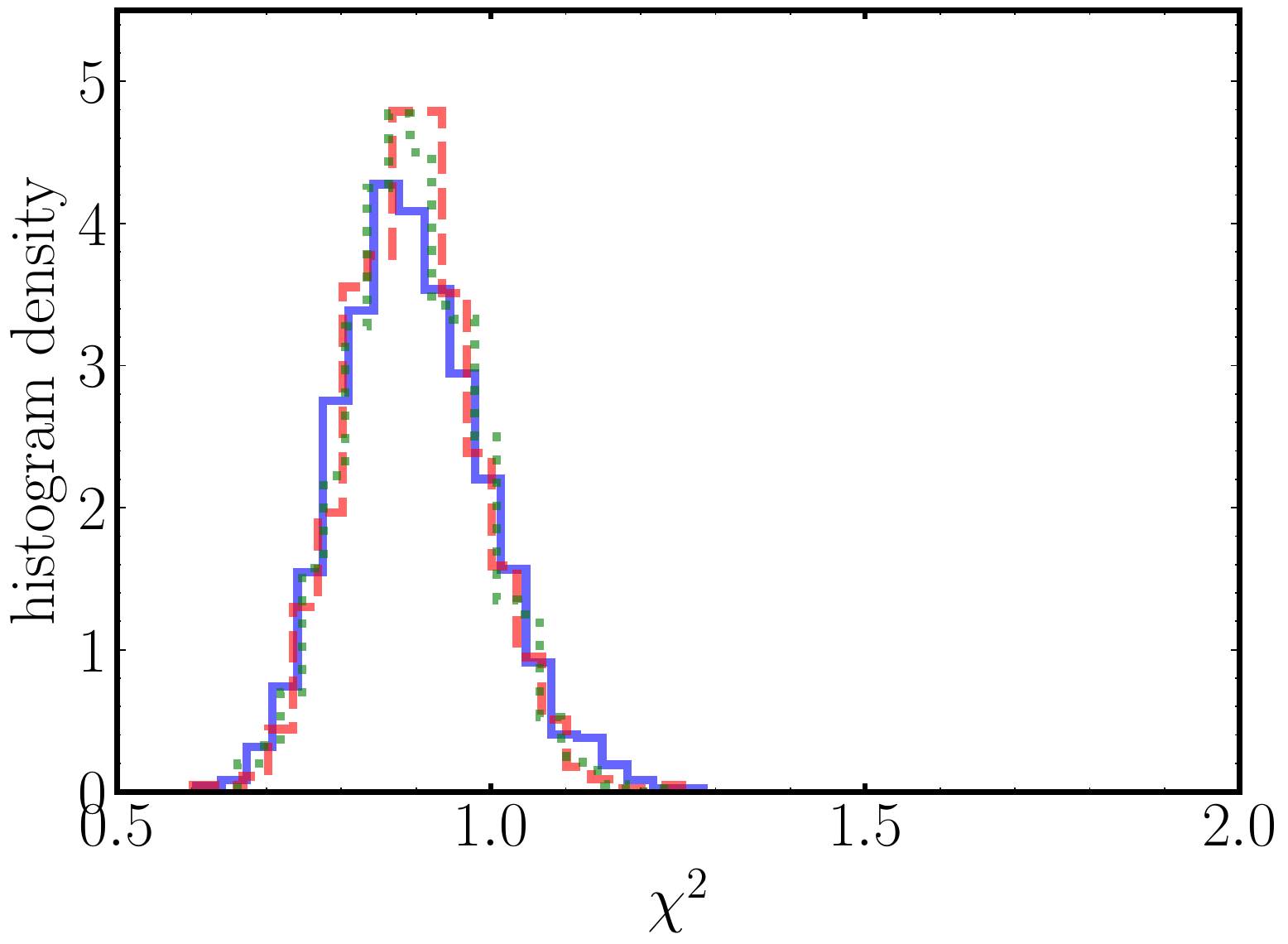}\\
\vspace{-0.8cm}\\
\hspace*{-0.28cm}\includegraphics[width=0.345\textwidth]{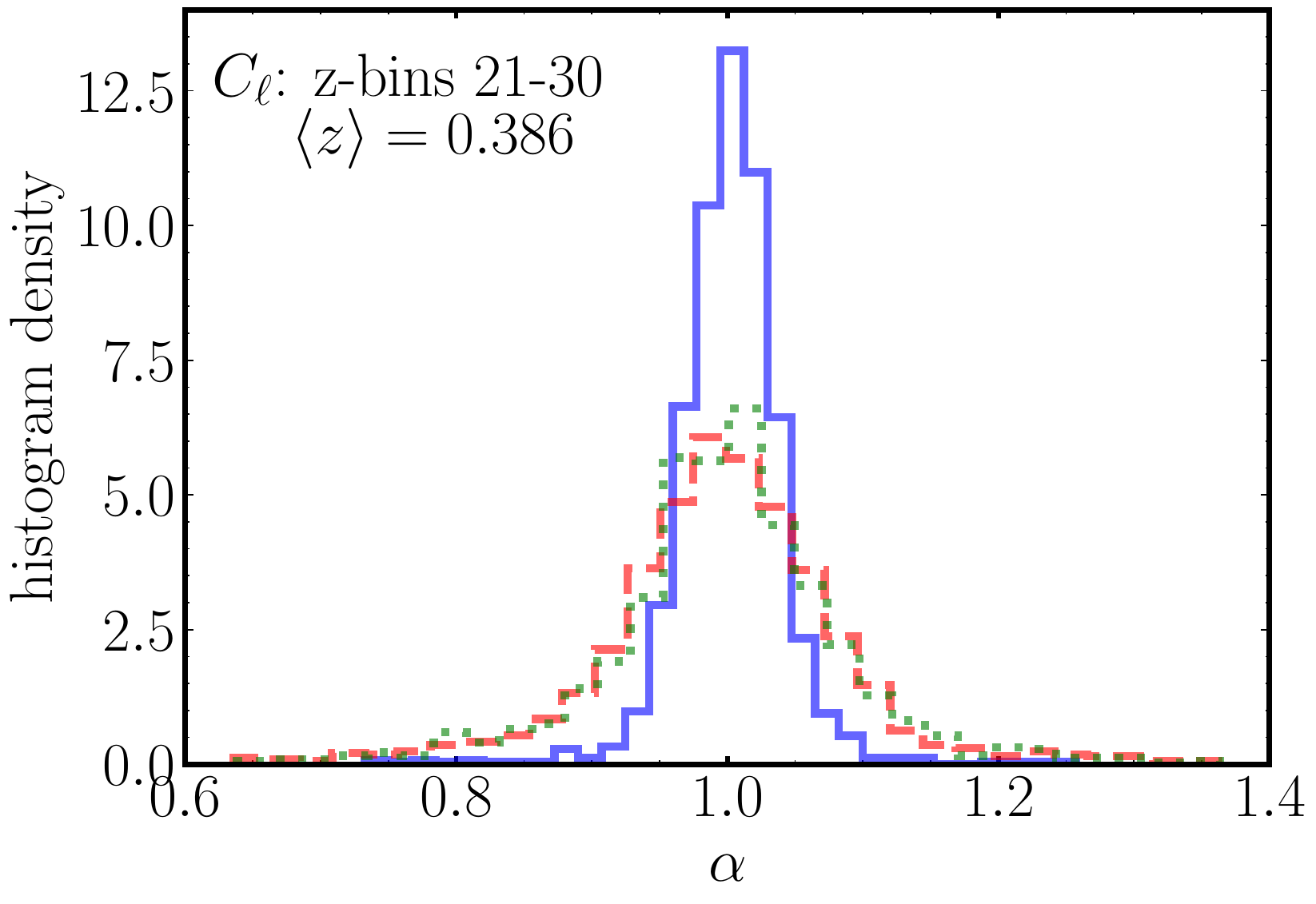}\includegraphics[width=0.33\textwidth]{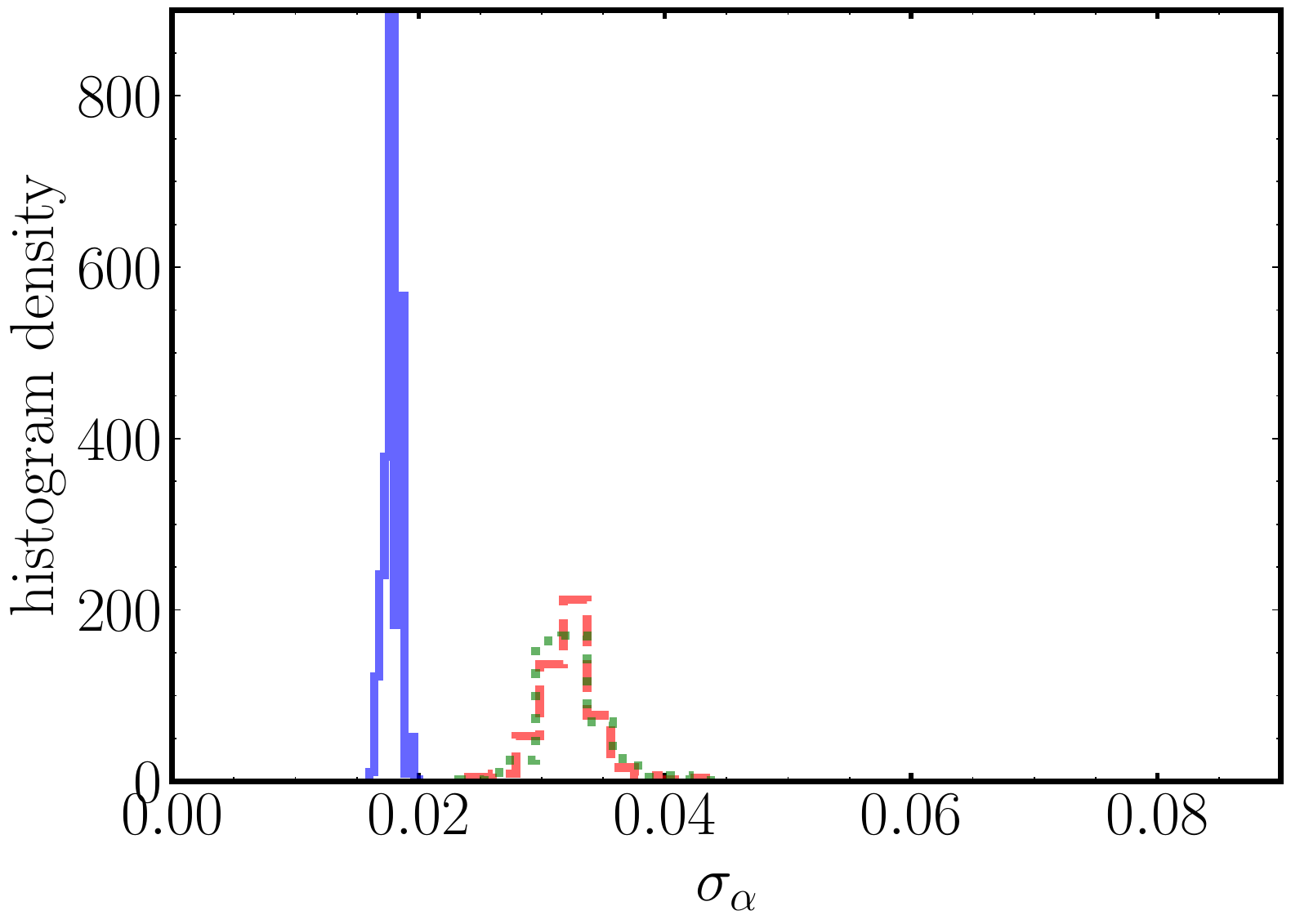}\hspace*{0.25cm}\includegraphics[width=0.33\textwidth]{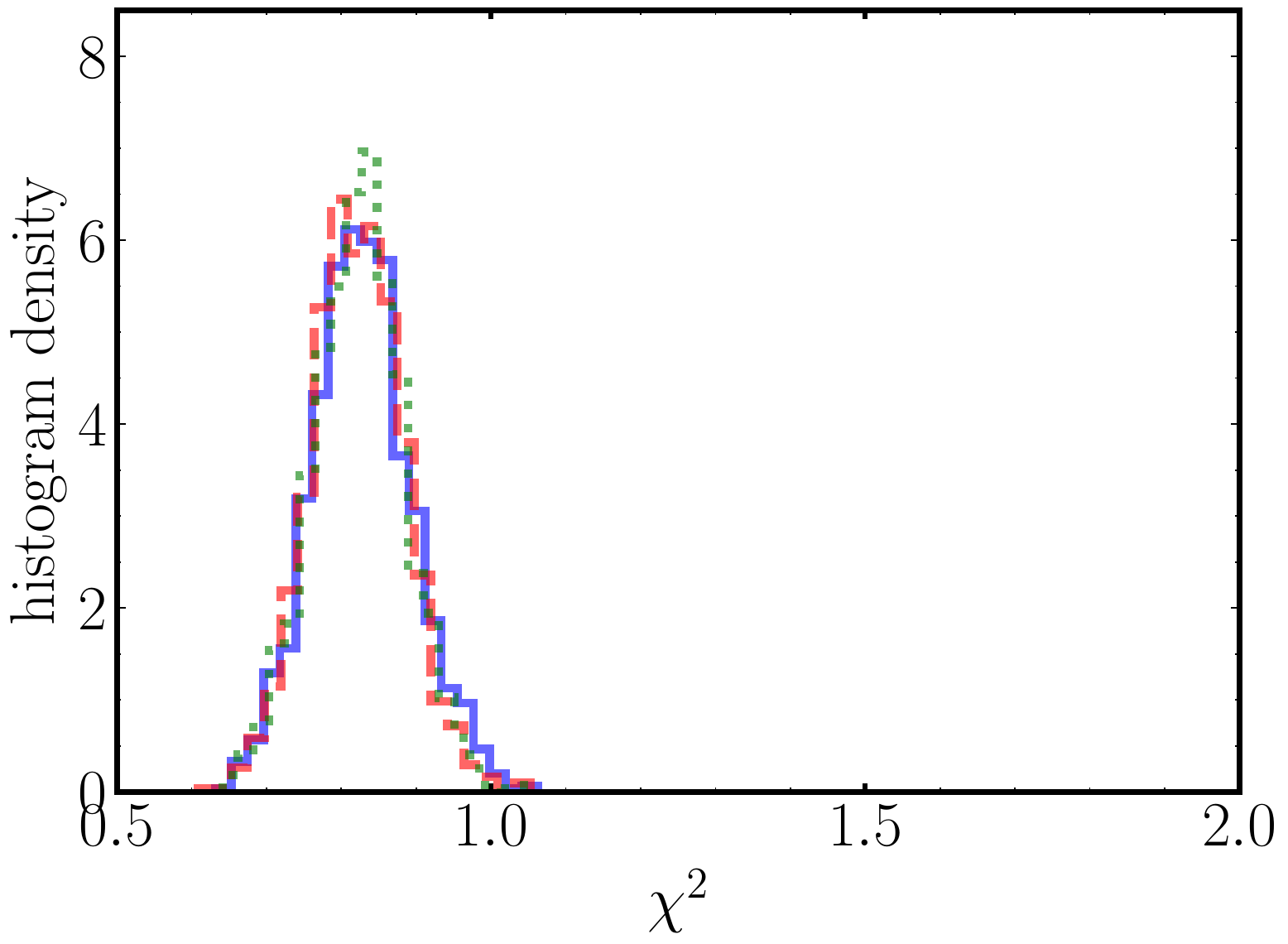}
\caption{Same as Fig. \ref{fig:baoResults-APS} but for the APS estimator. Again all distributions are constructed from the corresponding $N_s$ fraction of the mocks. See Table \ref{tab:results-21cmWNFG-APS}. }
\label{fig:baoResults-APS}
\end{figure*}

In addition, Tables \ref{tab:results-21cmWNFG-ACF} to \ref{tab:results-21cmWNFG-APS-mocks} present the same summary statistics obtained applying the BAO fitting pipeline to each mock for each of the three redshift ranges. 
In particular, there we show the average shift parameter $\langle \alpha \rangle$, the average error in $\alpha$ estimates $\langle \sigma_\alpha \rangle$, and two measures of the dispersion of the $\alpha$ distribution, the symmetric error around $\langle \alpha \rangle$ encompassing 68\% of the mocks (less sensitive to the tails) $\sigma_{68}$, and the common standard deviation of the distribution $\sigma_{std}$, as well as, the average $\langle \chi^2 \rangle$ (and degrees of freedom, dof).
Notice that all these quantities are calculated from the fraction $N_s$ of the total 1500 mocks, which correspond to the mocks fulfilling our selection criteria, namely, those whose $\alpha \pm \sigma_\alpha$ interval are inside the range $[0.6,1.4]$ and that have $\chi^2 > \chi^2_{nw}$, where $\chi^2_{nw}$ is obtained from fitting a non-BAO template model to the clustering estimates (see Section \ref{sec:significance}).
This fraction $N_s$ is also presented in the tables, as well as the fraction $N_d$ of the mock whose $\alpha$ estimate is inside the range $[0.8,1.2]$. 
We consider these as the mocks with a BAO detection. 
Figures \ref{fig:baoResults-ACF} and \ref{fig:baoResults-APS} show, for ACF and APS, respectively, the distribution of the $\alpha$ parameter, the $\sigma_\alpha$ error and the $\chi^2$, estimated from the $N_s$ fraction of the mocks; the blue curves show the 21\,cm only results.

It is worth mentioning that it is commonly employed in literature the detection criterion of having the full interval $\alpha \pm \sigma_\alpha$ inside the prior range $[0.8,1.2]$ \citep[e.g.,][]{2018/ata-boss-collab, 2018/chan}. 
However, one can observe from these analyses that the distribution of the $\alpha$ parameter in general seems to be more Gaussian than what we find in our analyses (left panels in Figs. \ref{fig:baoResults-ACF} and \ref{fig:baoResults-APS}). 
A possible reason for this is the smaller redshift we consider here, where the contribution from non-linear effects is stronger. 
However, results from \cite{2017/villaescusa}, investigating BAO detection from 21\,cm signal for the SKA case, for the redshift range $0.35<z<3.05$, also show $\alpha$ distributions with clear deviations from a perfect Gaussian (but, note the smaller number of simulations employed there, namely, 100). 
As pointed by \cite{2018/chan} and \cite{2021/abbott-des}, a natural consequence from having a (approximate) Gaussian distribution is a reasonable concordance among the three different error measurements, that is, $\langle \sigma_\alpha \rangle \sim \sigma_{68} \sim \sigma_{std}$. 
This is not our case, as can be seen from Tables \ref{tab:results-21cmWNFG-ACF} and \ref{tab:results-21cmWNFG-APS} (as well as from Tables \ref{tab:results-21cmWNFG-ACF-mocks} and \ref{tab:results-21cmWNFG-APS-mocks}), indicating that $\langle \sigma_\alpha \rangle$ is not meaningful or representative of the error in the $\alpha$ measurements for individual mock realizations (see also Figs. \ref{fig:baoResults-ACF} and \ref{fig:baoResults-APS}). 
Our results show that, compared to $\langle \sigma_\alpha \rangle$, the errors given by $\sigma_{68}$ are overestimated by $\sim $ 18\% to 33\%, from the smallest to the highest z-bins, when using the ACF, while underestimated by $\sim$ 50\% for APS. 
Also, comparing the expected $\sigma^m_\alpha$ obtained fitting the mean $C_\ell$ and $\omega(\theta)$ to the average $\langle \sigma_\alpha \rangle$, we find a reasonable agreement for intermediate and higher z-bins, but not for the lower z-bins, in particular for the ACF estimator. 
In addition, although $\sigma_{std}$ is in better agreement with $\sigma_{68}$, we still find non negligible differences among them, which confirms our $\alpha$ distributions as non-Gaussian (a larger $\sigma_{std}$ indicate the presence of non-Gaussian tails).  
Such reasons motivated our choice of using the $\alpha$ values, instead of $\alpha \pm \sigma_\alpha$ ranges, belonging to the interval $[0.8,1.2]$ as criterion for a BAO detection (defining the $N_d$ fraction), as well as our choice of using the 68\% spread of the $\alpha$ distributions, $\sigma_{68}$, as the representative error in our measurements. 

Comparing the results obtained for the three redshift ranges we confirm both clustering estimators to perform better at higher redshifts, as expected.
Remember that the lower the redshift the larger the BAO angular scale, $\theta_{BAO}$, which for the BINGO redshift range, $0.127 < z < 0.449$, corresponds to $ 18^\circ \gtrsim \theta_{BAO} \gtrsim 7^\circ $, for the fiducial cosmology. 
It means that, at the smallest redshifts the $\theta_{BAO}$ is larger than the $15^\circ$ stripe of the BINGO coverage, leading to worse statistics at such z-bins.
Figure \ref{fig:ACF_vs_APS} compares APS and ACF results obtained from 21\,cm only simulations (first row), showing the improvement in their performance for intermediate and higher redshifts, as well as a high correlation among their results.
In fact, comparing the performance from each of the clustering estimators we found that they show different sensitivities to specific redshift ranges.  
Our results show that $C_\ell$ measurements lead to smaller $\sigma_{68}$ errors for intermediate and higher redshift ranges, namely, 20\% and 25\% smaller $\sigma_{68}$ when compared to $\omega(\theta)$. 
Also, both $N_s$ and $N_d$ fractions of the mocks are larger for the APS at lower and intermediate redshifts, and comparable at higher redshifts. 
For the $N_d$ fraction the APS has $\sim$1.8 higher probability of detecting BAO at lower redshifts at 21\,cm only mocks compared to the ACF.

Finally, comparing the two estimators we find, in summary, an apparent better performance of the APS over the ACF. 
Such conclusion is mainly based on the following aspects of the ACF results: the slightly larger $\sigma_{68}$ uncertainties at intermediate and higher redshifts; the quite larger bias on the average $\langle \alpha \rangle$, especially at lower z-bins, corresponding to 0.95$\sigma_{68}$; and the smaller $N_s$ and $N_d$ fractions at small and intermediate z-bins.

Notice that we also evaluate the z-bin width and the way the 30 z-bins are combined to estimate the shift parameter. 
First, the BAO fitting pipeline was applied to three ranges of redshift composed by  $N_z = 12, 10, 8$ consecutive z-bins with $\delta\nu=9.33$\,MHz, so that $\alpha$ is estimated over three redshift ranges of same total width each ($\Delta z \approx 0.10$). 
Then, we test a tomographic binning of the BINGO frequency band into 30 z-bins linearly spaced in redshift ($\delta z \approx 0.01$), instead of frequency, but still estimating $\alpha$ for three redshift ranges of $N_z = 10$ z-bins each (same total redshift width). 
We find that, in general, both cases lead to results comparable to those from Tables \ref{tab:results-21cmWNFG-ACF} and \ref{tab:results-21cmWNFG-APS}, although each redshift range and clustering estimator has its own most advantageous configuration. 
In this sense, for simplicity, and because previous BINGO papers employ a redshift binning linear in frequency \citep{2021/liccardo_BINGO-sky-simulation, 2021/fornazier_BINGO-component_separation, 2021/costa_BINGO-forecast}, we keep using z-bins with $\delta \nu = 9.33$\,MHz width and the BAO fitting process considering three sets of $N_z=10$ z-bin for both estimators.

\begin{table*}
\linespread{1.4}
\selectfont
\centering
\caption{BAO fitting results for the ACF using 1500 {\tt FLASK} mocks. The three first parts of the table correspond to analyses of 21\,cm only, noisy 21\,cm (+ white noise) and BINGO-like (+ white noise + foreground residuals) mocks, using in both cases the fiducial configuration. The other parts of the table show results from the robustness tests, where the first column points out the aspects varying with respect to the fiducial configuration, namely the $\Delta \theta$ binning width and the $q$ index from $\omega(\theta)$ template. All robustness tests performed over the BINGO-like mocks. The last part of the table corresponds to analyses of the BINGO-like mocks, employing the fiducial configuration, when varying the beam size with frequency. The last column shows $\alpha^m \pm \sigma^m_{\alpha}$ obtained from fitting the mean clustering from the mocks. See text for details.}
{
\begin{tabular}{|p{0.12\textwidth}|c|c|c|c|c|c|c|c|c|c|}
\hline
Tests \,\,& z-bins & $\langle \alpha \rangle$ & $\langle \sigma_\alpha \rangle$ & $\sigma_{68}$ & $\sigma_{std}$ & $\langle \chi^2 \rangle$/dof ($\chi^2_{red}$) & $N_s$ (\%) & $N_d$ (\%) & Mean \\
\hline \hline
\multicolumn{10}{|c|}{21\,cm only} \\
\hline
\multirow{3}{*}{\makecell{Fid. config.}} &  1--10 & 1.1158 & 0.1427 & 0.1209 & 0.1166 & 122.49/123 (1.00) & 65.27 & 45.73 & 1.0067 $\pm$ 0.2433 \\
                 & 11--20 & 1.0426 & 0.0977 & 0.0775 & 0.0923 &  76.84/70 (1.10) & 81.00 & 75.00 & 1.0082 $\pm$ 0.1014 \\
                 & 21--30 & 1.0060 & 0.0565 & 0.0425 & 0.0514 &  69.54/61 (1.14) & 92.60 & 92.00 & 1.0076 $\pm$ 0.0507 \\
\hline \hline
\multicolumn{10}{|c|}{+ white noise} \\
\hline
                             &  1--10 & 1.0906 & 0.1619 & 0.1042 & 0.1066 & 122.05/123 (0.99) & 58.73 & 48.60 & 1.0070 $\pm$ 0.2915 \\
Fid. Config.                   & 11--20 & 1.0524 & 0.1213 & 0.0924 & 0.0948 &  76.34/70 (1.09)  & 80.53 & 73.80 & 1.0107 $\pm$ 0.1470 \\
                             & 21--30 & 1.0267 & 0.0932 & 0.0807 & 0.0883 &  67.88/61 (1.11)  & 89.60 & 85.13 & 1.0070 $\pm$ 0.0989 \\
                             \hline \hline
\multicolumn{10}{|c|}{+ white noise + foreground residuals} \\
\hline
\multirow{3}{*}{\makecell{Fid. Config.\\ ($\Delta \theta = 0.50^\circ$; \\$q = 0, 1, 2$)}} &  1--10 & 1.0897 & 0.1624 & 0.1123 & 0.1118 & 122.42/123 (1.00) & 57.07 & 46.27 & 0.9952 $\pm$ 0.3270 \\
                             & 11--20 & 1.0362 & 0.1267 & 0.0969 & 0.1070 & 76.60/70 (1.09) & 70.20 & 63.93 &  0.9989 $\pm$ 0.1546 \\
                             & 21--30 & 1.0074 & 0.1006 & 0.0809 & 0.0930 & 68.37/61 (1.12) & 78.00 & 73.53 & 0.9965 $\pm$ 0.1039 \\
\hline \hline
                             &  1--10 & 1.0901 & 0.1649 & 0.1018 & 0.1080 & 238.66/287 (0.83) & 53.40 & 44.67 & 0.9670 $\pm$ 0.6540 \\
$\Delta \theta = 0.25^\circ$ & 11--20 & 1.0370 & 0.1307 & 0.0999 & 0.1054 & 166.28/180 (0.92) & 71.67 & 64.87 & 1.0135 $\pm$ 0.2560 \\
                             & 21--30 & 1.0063 & 0.1034 & 0.0842 & 0.0970 & 154.15/164 (0.94) & 76.67 & 72.33 & 0.9826 $\pm$ 0.1394 \\
\hline
                             &  1--10 & 1.0888 & 0.1647 & 0.1100 & 0.1157 & 155.02/165 (0.94) & 57.73 & 46.60 & 1.0034 $\pm$ 0.3245 \\
$\Delta \theta = 0.40^\circ$ & 11--20 & 1.0375 & 0.1298 & 0.0969 & 0.1054 &  100.85/98 (1.03) & 72.00 & 65.20 & 0.9981 $\pm$ 0.1597 \\ 
                             & 21--30 & 1.0160 & 0.0998 & 0.0833 & 0.0982 &  90.00/86 (1.05)  & 78.80 & 73.07 & 0.9976 $\pm$ 0.1065 \\
\hline \hline
                             &  1--10 & 1.1687 & 0.1299 & 0.1092 & 0.1117 & 130.45/133 (0.98) & 37.40 & 18.73 & 0.9894 $\pm$ 0.3321 \\
$q = 0, 1$                   & 11--20 & 1.0364 & 0.1294 & 0.1022 & 0.1070 &  85.43/80 (1.07)  & 70.00 & 63.60 &  0.9955 $\pm$ 0.1521 \\
                             & 21--30 & 1.0068 & 0.1005 & 0.0823 & 0.0951 &  78.36/71 (1.10)  & 85.80 & 80.67 & 1.0023 $\pm$ 0.0989 \\
\hline
                             &  1--10 & 1.0790 & 0.1813 & 0.1432 & 0.1269 & 114.55/113 (1.01) & 34.93 & 26.47 & 1.0015 $\pm$ 0.3270 \\
$q~=~-1,~0,~1,~2$            & 11--20 & 1.0344 & 0.1307 & 0.0963 & 0.1018 &  67.38/60 (1.12)  & 70.53 & 65.20 & 1.0047 $\pm$ 0.1521 \\
                             & 21--30 & 1.0200 & 0.0954 & 0.0801 & 0.0895 &  58.92/51 (1.14)  & 85.87 & 82.20 & 1.0096 $\pm$ 0.1014\\
\hline
                             &  1--10 & 1.0995 & 0.1657 & 0.1046 & 0.1039 & 216.71/123 (1.76) & 14.20 & 11.33 & 0.9952 $\pm$ 0.7275 \\
$q = -2, 0, 1$               & 11--20 & 1.0444 & 0.1214 & 0.0839 & 0.0904 &  146.87/70 (2.10) & 39.87 & 37.73 & 0.9989 $\pm$ 0.1648 \\
                             & 21--30 & 1.0178 & 0.0963 & 0.0790 & 0.0891 &  101.07/61 (1.67) & 63.87 & 61.07 & 0.9965 $\pm$ 0.1065 \\
                             \hline \hline
\multicolumn{10}{|c|}{+ white noise + foreground residual (varying beam size)} \\
\hline
\multirow{3}{*}{\makecell{Fid. config.}}
                             & 1--10  & 1.0929 & 0.1620 & 0.1162 & 0.1157 & 121.86/123 (0.99) & 58.20 & 46.33 & 0.9994 $\pm$ 0.3118 \\
                             & 11--20 & 1.0419 & 0.1245 & 0.1003 & 0.1113 &   77.35/70 (1.10) & 72.87 & 64.27 & 1.0004 $\pm$ 0.1546 \\
                             & 21--30 & 1.0164 & 0.0900 & 0.0665 & 0.0871 &   72.09/61 (1.18) & 80.53 & 76.27 & 1.0030 $\pm$ 0.0887 \\
\hline
\end{tabular}}
\label{tab:results-21cmWNFG-ACF}
\end{table*}

\begin{table*}
\linespread{1.4}
\selectfont
\centering
\caption{Same as Table \ref{tab:results-21cmWNFG-ACF} but for the APS estimator.}
{
\begin{tabular}{|p{0.12\textwidth}|c|c|c|c|c|c|c|c|c|c|}
\hline
Tests \,\,& z-bins & $\langle \alpha \rangle$ & $\langle \sigma_\alpha \rangle$ & $\sigma_{68}$ & $\sigma_{std}$ & $\langle \chi^2 \rangle$/dof &  $N_s$ (\%) & $N_d$ (\%) &  Mean  \\
\hline \hline
\multicolumn{10}{|c|}{21\,cm only} \\
\hline
\multirow{3}{*}{\makecell{Fid. config.}} &  1--10 & 1.0061 & 0.0594 & 0.1200 & 0.1243 & 102.14/100 (1.02) & 93.73 & 84.20 & 1.0012 $\pm$ 0.0608 \\
                 & 11--20 & 1.0083 & 0.0316 & 0.0616 & 0.0889 & 185.70/207 (0.90) & 92.87 & 87.73 & 1.0062 $\pm$ 0.0304 \\
                 & 21--30 & 1.0024 & 0.0178 & 0.0318 & 0.0399 & 244.17/295 (0.83) & 92.93 & 92.27 & 1.0028 $\pm$ 0.0152 \\
\hline \hline
\multicolumn{10}{|c|}{+ white noise} \\
\hline
\multirow{3}{*}{\makecell{Fid. config.}} &  1--10 & 1.0041 & 0.0653 & 0.1311 & 0.1311 &  102.12/100  (1.02) & 92.33 & 79.20 & 0.9915 $\pm$ 0.0684 \\
                   & 11--20 & 1.0057 & 0.0410 & 0.0929 & 0.1092 &  185.46/207 (0.91)  & 90.87 & 83.27 & 1.0036 $\pm$ 0.0406 \\
                   & 21--30 & 0.9965 & 0.0322 & 0.0707 & 0.0897 &  242.42/295 (0.82)  & 91.60 & 86.87 & 0.9999 $\pm$ 0.0304 \\
                             \hline \hline
\multicolumn{10}{|c|}{+ white noise + foreground residual} \\
\hline
\multirow{3}{*}{\makecell{Fid. Config.\\ ($\Delta \ell = 10$; \\ $q = -1, 0, 1, 2$)}} &  1--10 & 1.0009 & 0.0659 & 0.1257 & 0.1274 & 102.23/100 (1.02) & 90.47 & 78.87 & 0.9925 $\pm$ 0.0684 \\
                   & 11--20 & 1.0012 & 0.0418 & 0.0888 & 0.1063 & 185.53/207 (0.90) & 92.47 & 84.77 & 0.9985 $\pm$ 0.0431 \\
                   & 21--30 & 0.9975 & 0.0323 & 0.0701 & 0.0890 & 242.88/295 (0.82) & 88.13 & 83.73 & 1.0012 $\pm$ 0.0304 \\
\hline \hline
                   &  1--10 & 1.0253 & 0.0697 & 0.1226 & 0.1205 &  45.23/36 (1.26)  & 94.07 & 82.67 & 1.0182 $\pm$ 0.0710 \\
$\Delta \ell = 15$ & 11--20 & 1.0070 & 0.0474 & 0.1002 & 0.1154 & 108.96/108 (1.01) & 92.87 & 83.00 & 1.0016 $\pm$ 0.0482 \\ 
                   & 21--30 & 1.0015 & 0.0345 & 0.0685 & 0.0867 & 157.06/168 (0.93) & 94.47 & 90.40 & 1.0020 $\pm$ 0.0355 \\
\hline
                   &  1--10 & 1.0299 & 0.0724 & 0.1293 & 0.1128 &  27.24/17 (1.60)  & 96.87 & 87.80 & 1.0346 $\pm$ 0.0760 \\
$\Delta \ell = 20$ & 11--20 & 1.0199 & 0.0476 & 0.1088 & 0.1117 &  76.90/71 (1.08)  & 94.60 & 86.93 & 1.0117 $\pm$ 0.0456 \\
                   & 21--30 & 1.0009 & 0.0366 & 0.0726 & 0.0912 & 116.66/117 (1.00) & 96.20 & 91.47 & 0.9975 $\pm$ 0.0355 \\
\hline \hline
                   &  1--10 & 1.0523 & 0.0682 & 0.1307 & 0.1334 & 119.73/120 (1.00) & 62.73 & 50.80 & 1.0985 $\pm$ 0.0735 \\
$q = 0, 1$         & 11--20 & 1.0558 & 0.0436 & 0.0788 & 0.0870 & 199.91/227 (0.88) & 73.33 & 68.47 & 1.0626 $\pm$ 0.0431 \\
                   & 21--30 & 1.0279 & 0.0333 & 0.0517 & 0.0556 & 254.90/315 (0.81) & 76.27 & 75.67 & 1.0294 $\pm$ 0.0330 \\
\hline
                   &  1--10 & 0.9995 & 0.0660 & 0.1236 & 0.1279 & 110.52/110 (1.00) & 71.87 & 62.40 & 0.9935 $\pm$ 0.0684 \\
$q = 0, 1, 2$      & 11--20 & 0.9912 & 0.0419 & 0.0858 & 0.1091 & 192.63/217 (0.89) & 82.73 & 74.80 &  0.9960 $\pm$ 0.0406 \\
                   & 21--30 & 0.9958 & 0.0324 & 0.0637 & 0.0818 & 248.84/305 (0.82) & 84.33 & 81.53 & 0.9987 $\pm$ 0.0304 \\
\hline
                   &  1--10 & 1.0161 & 0.0664 & 0.1324 & 0.1278 & 110.95/110 (1.01) & 83.73 & 72.40 & 1.0113 $\pm$ 0.0684 \\
$q = -2, 0, 1$     & 11--20 & 1.0162 & 0.0422 & 0.0864 & 0.1020 & 193.18/217 (0.89) & 88.33 & 82.13 & 1.0164 $\pm$ 0.0431 \\
                   & 21--30 & 1.0097 & 0.0326 & 0.0691 & 0.0870 & 249.10/305 (0.82) & 84.47 & 80.27 & 1.0131 $\pm$ 0.0304 \\
                             \hline \hline
\multicolumn{10}{|c|}{+ white noise + foreground residual (varying beam size)} \\
\hline
\multirow{3}{*}{\makecell{Fid. config.}} & 1--10  & 1.0037 & 0.0642 & 0.1315 & 0.1305 & 102.23/100 (1.02) & 91.87 & 79.00 & 0.9946 $\pm$ 0.0659 \\
                   & 11--20 & 0.9984 & 0.0408 & 0.0922 & 0.1088 & 185.68/207 (0.90) & 92.13 & 83.67 & 0.9975 $\pm$ 0.0406 \\
                   & 21--30 & 0.9860 & 0.0338 & 0.0524 & 0.0670 & 337.97/295 (1.15) & 75.87 & 73.93 & 0.9939 $\pm$ 0.0330 \\
\hline
\end{tabular}}
\label{tab:results-21cmWNFG-APS}
\end{table*}

\subsection{Impact of including instrumental noise} \label{sec:results-flask-noise}

Taking into account all specifications summarized in Section \ref{sec:foreg}, we estimate the white noise level at each pixel in the BINGO region, $\sigma_{pix}$, considering the same resolution used to generate the 21\,cm simulations, $N_{side} = 256$. 
From this $\sigma_{pix}$ map we can generate as many realizations of the corresponding white noise map as necessary by multiplying it by random values defined by a Gaussian distribution of zero mean and unitary standard deviation. 
We add a different noise realization to each of the 1500 {\tt FLASK} mocks and repeat the same process of calculating $C_\ell$ and $\omega(\theta)$ from each of them. 

White noise adds to the APS a constant term $N_\ell$ at all multipoles, which for BINGO dominates over the 21\,cm signal, as a function of redshift bin, for $\ell \geq 200-350$ (Fig. \ref{fig:Cl-Nl}). 
Then, we debias the $C_\ell$ of each noisy 21\,cm simulation by subtracting from it the expected $N_\ell$ amplitude, computed from the theoretical noise level.
In contrast, white noise affects the ACF over all angular scales, so that this debiasing cannot be considered. 
However, since the  $\omega(\theta)$ is calculated by averaging over $\delta T_p \delta T_{p'}$ from all pairs of pixels $p$ and $p'$ separated by $\theta$ (Eq. (\ref{eq:ACF-temp})), a natural consequence is a reduction of the noise. 

Our results show that the presence of thermal noise increase the $\sigma_{68}$ uncertainty by 19\% and 90\% for intermediate and higher z-bins when using the ACF, while at lower z-bins this error surprisingly diminishes by 14\% (similarly to the respective bias on $\langle \alpha \rangle$). 
The noise impact is more expressive for APS analyses, with 9\%, 51\% and 122\% increase in $\sigma_{68}$ from lower to higher z-bins. 
However, although APS results are more affected by the noise, it still has a smaller $\sigma_{68}$ uncertainty for higher z-bins, while for intermediate z-bins the two estimators have comparable errors. 
The second row of Fig. \ref{fig:ACF_vs_APS} shows a comparison of the best-fit $\alpha$ from $\omega(\theta)$ and $C_\ell$ estimators,  clearly much more spread due to the presence of noise.
In addition, for both estimators the $N_s$ and $N_d$ fractions decrease due to the presence of thermal noise, with the greatest impact over $N_s$, decreasing by 10\% for the ACF estimator at lower z-bins. 
Again surprising, $N_d$ increases by 6\% for ACF at small z-bins after including noise. 
Apart from this, the number of simulations with BAO detection has the larger decrease the higher the redshift. 
In summary, the presence of thermal noise has the major impact on higher z-bins, for which the BAO feature appears at smaller angular scales (spread to larger multipoles; Fig. \ref{fig:Cl-Nl}), where noise starts dominating.

\subsection{Foreground residual contamination} \label{sec:results-flask-foreg}

The last step is to take into account the presence of residual foreground as expected for BINGO. 
For this we add, along with the thermal noise, the most important foreground signals contributing to the BINGO frequency range (Section \ref{sec:foreg}) to a 21\,cm realization. 
Using the {\tt GNILC} code, we perform component separation to reconstruct the noisy 21\,cm signal as discussed in Section \ref{sec:gnilc}. 
Also, from Eq. (\ref{eq:recHI_plus_residue}), using the ILC filter and the known foreground contamination we can estimate the amplitude of the foreground residual in each z-bin, $\mathbf{W f}$. 
Such process is repeated for 10 different realizations of the 21\,cm signal, each of them contaminated by a different realization of the thermal noise, but all with the same foreground signal contribution. 
These 10 different estimates of the foreground residual is used to include the expected contaminant signal to each of the 1500 mock realizations (the BINGO-like simulations). 
Note that the noise realizations are still different from one mock to another, while the foreground residual will be repeated for each 10 mocks. 
This avoids the (computationally expensive) component separation process to be applied to all the simulations.

\begin{figure*}[h]
\centering
\includegraphics[width=0.8\textwidth]{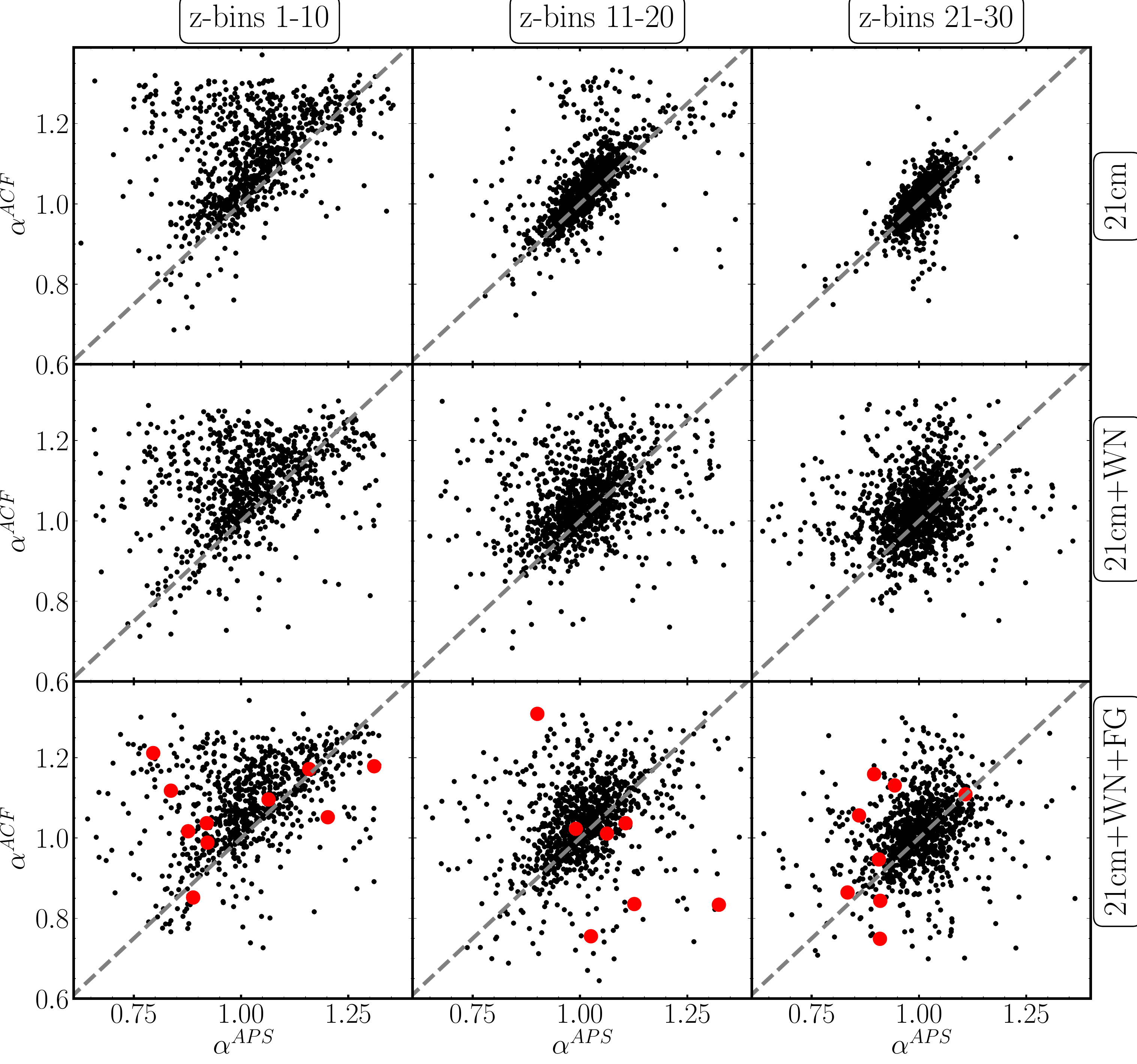}
\caption{Comparison of the $\alpha$ values estimated using the ACF and APS estimators (black dots). The columns from left to right show estimates from lower to higher redshift ranges, with average $\langle z \rangle = 0.172$, $\langle z \rangle = 0.270$, $\langle z \rangle = 0.386$. The first row shows results for the 21\,cm only analyses, while the middle and bottom rows show results including white noise only and white noise + foreground residual, respectively. The red dots on the bottom row show $\alpha$ estimates from the reconstructed (foreground cleaned) maps. }
\label{fig:ACF_vs_APS}
\end{figure*}

Given the 1500 semi-realistic BINGO-like simulations, we calculate the $C_\ell$ and $\omega(\theta)$ from each of them and repeat the BAO fitting process using the fiducial configuration. 
Note that, again we debias the $C_\ell$ by subtracting the expected $N_\ell$ term. 
The results are shown in Tables \ref{tab:results-21cmWNFG-ACF} and \ref{tab:results-21cmWNFG-APS}, as well as in Figs. \ref{fig:baoResults-ACF} and \ref{fig:baoResults-APS}, for ACF and APS estimators, respectively. 
The last row of Fig. \ref{fig:ACF_vs_APS} also shows how the foreground residual can affect the concordance among the two estimators.  
In general, one can see that such contamination has less significant impact, with negligible changes on both $\langle \sigma_\alpha \rangle$, $\sigma_{68}$ and $\sigma_{std}$ uncertainties compared to including noise \citep[similar conclusions are drawn by][]{2017/villaescusa}.
Still, both $N_s$ and $N_d$ fractions decrease, especially at higher z-bins using ACF, with another $\sim$ 10\% of the 1500 mocks no longer satisfying the selection and detection criteria. 
Note also that, for all the redshift ranges, the best-fit $\alpha^m$ from the mean ACF shows a slight shift to values smaller than those found including only noise. 
Since this shift persists at some redshift ranges when changing the fiducial configuration (Table \ref{tab:results-21cmWNFG-ACF}) and also when using the N-body mocks (Table \ref{tab:results-21cmWNFG-ACF-mocks}), it suggests the foreground residual as the possible responsible for such systematic bias. 
For the APS, on the other hand, the results do not lead to the same conclusion. 
In any case, even though such shift is observed from $\omega(\theta)$ results after including the foreground residual contribution, it is quite small with respect to the error estimates, that is, small enough to not be a concern. 

In order to evaluate the validity of including the expected foreground residual to the mocks, instead of applying the {\tt GNILC} code to each of them, we add to the last row of Fig. \ref{fig:ACF_vs_APS} the BAO fitting results from the 10 reconstructed maps.
As one can see, the red dots appear to be in good agreement with the black ones. 
The middle and last panels do not show all 10 red dots because not all the reconstructed maps provided best-fit $\alpha$ values in the range $[0.6,1.4]$, for (at least) one of the estimators.

Additionally, we test the impact of a frequency dependent beam size, with the BINGO $\theta_{\mbox{\sc fwhm}}$ taking values from $\sim\,35'$ to $\sim\,45'$ from small to high z-bins. 
In this case, the component separation is applied after each frequency map (21\,cm, along with thermal noise and foreground signals) has its resolution converted to the common $\theta_{\mbox{\sc fwhm}} \sim 45'$, the largest beam size. 
The foreground residual for each z-bin is estimated as before. 
Note that, the foreground residual is added to a realization after each frequency map, containing the 21\,cm signal ({\tt FLASK} mocks), convolved with the corresponding beam size, and the thermal noise, is converted to the common $\theta_{\mbox{\sc fwhm}}$. 
The BAO fitting results from this set of 1500 mock realizations are presented in the last part of Tables \ref{tab:results-21cmWNFG-ACF} and \ref{tab:results-21cmWNFG-APS}. 
These results show that, for both clustering estimators, a frequency dependent beam size have negligible impact on the BAO analyses for lower and intermediate redshift ranges. 
At higher z-bins one can observe a slightly larger bias on $\langle \alpha \rangle$ for both, $\omega(\theta)$ and $C_\ell$, but still quite smaller than the error estimates, besides a decrease and slight increase in both $N_s$ and $N_d$ fractions for $C_\ell$ and $\omega(\theta)$, respectively. 
In fact, we find that the foreground residual have a larger amplitude on the higher z-bins when compared to the case with a fixed beam size (Figure \ref{fig:Cl-Nl}), contributing mainly at larger multipoles (a deeper investigation on the impact of frequency dependent beam size in the foreground cleaning is necessary but out of the scope of this paper and will be considered in future work). 
Still, even with larger contribution, the foreground residual does not seem to have a significant impact on BAO results, allowing us to keep the previous conclusions.

\subsection{Robustness tests} \label{sec:robustness-tests}

Here we test the aspects of the fiducial configuration, that is, we consider variations of (a) the angular ($\Delta\theta$) and multipole ($\Delta\ell$) binning and (b) the $q$ index in the templates, Eqs. (\ref{eq:APS-temp}) and (\ref{eq:ACF-temp}). 
Such robustness tests are performed over the BINGO-like simulations because they are the closest to what BINGO will observe. 
Notice that we vary only one aspect of the fiducial configuration at a time, the one pointed in the first column of Tables \ref{tab:results-21cmWNFG-ACF} and \ref{tab:results-21cmWNFG-APS}, where the results of all robustness tests are summarized. 

From testing $\Delta \theta$ width, Table \ref{tab:results-21cmWNFG-ACF} shows that the different binning lead to small changes at distinct statistics. 
The most discrepant result concerns the larger bias on $\alpha^m$ obtained from the mean $\omega(\theta)$, at all the redshift ranges, for $\Delta\theta = 0.25^\circ$. 
Also, although $\Delta \theta = 0.40^\circ$ shows slightly better results for some statistics, for example, larger $N_s$ and $N_d$ fractions at some redshift ranges, we find a smaller bias on $\langle \alpha \rangle$ at higher z-bins for $\Delta \theta = 0.50^\circ$. 
In this sense, and in order to have a more diagonal covariance matrix, we choose to use $\Delta \theta = 0.50^\circ$ as the fiducial configuration. 
Similar conclusions can be drawn from the tests on the $\omega(\theta)$ template, showing that some choices of $q$ can help improving different statistics. 
However, by comparing all template tests, one can notice the large reduced $\langle \chi^2 \rangle$, for all redshift ranges, using $q = -2, 0, 1$, and the large bias on $\langle \alpha \rangle$ and $\alpha^m$ for the ACF at small z-bins for $q = 0, 1$. 
In general, we see that cases $q = 0, 1, 2$ and $q = -1, 0, 1, 2$ present the better results, especially in terms of the fraction $N_d$ of mocks with BAO detection. 
Indeed, the former one produces larger $N_s$ and $N_d$ for lower z-bins, while the latter provides larger fractions at higher z-bins.  
Given the smaller bias amplitude in $\langle \alpha \rangle$ at higher z-bins, we choose $q = 0, 1, 2$ as part of the fiducial configuration.

Results from testing $\Delta \ell$ width, presented in Table \ref{tab:results-21cmWNFG-APS}, show a good concordance between the different uncertainties from all three cases considered, although a clear worse performance for $\Delta \ell = 20$.
While this larger bin width gives larger $N_s$ and $N_d$ fractions, it also introduces large bias on $\langle \alpha \rangle$ and $\alpha^m$ for most of the redshift ranges. 
Comparing $\Delta \ell = 10$ and $15$, the second one leads to slightly larger $N_s$ and $N_d$ fractions, but also a larger bias on $\langle \alpha \rangle$ and $\alpha^m$, as well as a larger $\langle \chi^2 \rangle$, for the lower z-bins. 
Then, we find more appropriate to use $\Delta \ell = 10$ as part of the fiducial configuration. 
Finally, testing the different templates to fit $C_\ell$ measurements, we look again for the one with the smallest bias and uncertainty, combined with the largest $N_s$ and $N_d$. 
In this sense, one can see that $q = 0,1,2$ and $q = -1, 0, 1, 2$ have the best results, but the latter gives a larger fraction of the mocks with BAO detection, $N_d$, and also larger $N_s$, motivating the choice of using it as part of the fiducial configuration. 

From all the robustness tests, considering each redshift range individually, one could say that each of them would require a slightly different configuration. 
Still, just to make things simpler, we chose to use the same $\theta$/$\ell$ binning and $\omega(\theta)$/$C_\ell$ templates for all three redshift ranges, that is, the (a) and (b) aspects of the fiducial configuration, as discussed.

\begin{table*}
\linespread{1.3}
\selectfont
\centering
\caption{BAO fitting results for the ACF using 500 N-body mocks. The first part of the table show results obtained analyzing 21\,cm only mock, while the second part correspond to BINGO-like simulations. In both cases the fiducial configuration is employed.}
{
\begin{tabular}{|c|c|c|c|c|c|c|c|c|c|}
\hline
Tests \,\,& z-bins & $\langle \alpha \rangle$ & $\langle \sigma_\alpha \rangle$ & $\sigma_{68}$ & $\sigma_{std}$ & $\langle \chi^2 \rangle$/dof & $N_s$ (\%) & $N_d$ (\%)& Mean \\
                             \hline \hline
\multirow{3}{*}{\makecell{21\,cm only}}   &  1--10 & 1.1339 & 0.1411 & 0.1178 & 0.1085 & 98.31/123 (0.80) & 49.60 & 34.80 & 1.0105 $\pm$ 0.2991 \\
                 & 11--20 & 1.0354 & 0.0973 & 0.0958 & 0.1001 & 67.63/70 (0.97) & 79.40 & 73.40 & 1.0045 $\pm$ 0.1217 \\
                 & 21--30 & 1.0091 & 0.0560 & 0.0554 & 0.0642 & 61.79/61 (1.01) & 87.40 & 86.40 & 1.0035 $\pm$ 0.0558 \\
                             \hline \hline
\multirow{3}{*}{\makecell{+ white noise \\ + foreground \\ residual}}  &  1--10 & 1.1082 & 0.1605 & 0.1014 & 0.1074 & 97.59/123 (0.79) & 46.00    & 36.40 & 1.0194 $\pm$ 0.3752 \\
                 & 11--20 & 1.0424 & 0.1317 & 0.1082 & 0.1106 & 66.97/70 (0.96) & 68.00 & 59.60 & 0.9903 $\pm$ 0.1749 \\
                 & 21--30 & 0.9974 & 0.1024 & 0.1160 & 0.1209 & 61.04/61 (1.00) & 75.60 & 66.20 & 0.9698 $\pm$ 0.1242 \\
\hline 
\end{tabular}}
\label{tab:results-21cmWNFG-ACF-mocks}
\end{table*}

\begin{table*}
\linespread{1.3}
\selectfont
\centering
\caption{Same as Fig. \ref{tab:results-21cmWNFG-ACF-mocks} but for the APS estimator. }
{
\begin{tabular}{|c|c|c|c|c|c|c|c|c|c|}
\hline
Tests \,\,& z-bins & $\langle \alpha \rangle$ & $\langle \sigma_\alpha \rangle$ & $\sigma_{68}$ & $\sigma_{std}$ & $\langle \chi^2 \rangle$/dof & $N_s$ (\%) & $N_d$ (\%)& Mean \\
                             \hline \hline
\multirow{3}{*}{\makecell{21\,cm only}}   &  1--10 & 1.0075 & 0.0385 & 0.1373 & 0.1402 & 84.81/100 (0.85) & 84.00 & 68.60    & 1.0110 $\pm$ 0.0380 \\
                 & 11--20 & 1.0128 & 0.0190 & 0.0682 & 0.0873 & 119.81/207 (0.58) & 78.40 & 74.60 &  1.0062 $\pm$ 0.0177 \\
                 & 21--30 & 1.0096 & 0.0130 & 0.0359 & 0.0549 & 117.95/295 (0.40) & 78.20 & 76.60 &  1.0086 $\pm$ 0.0127 \\
                             \hline \hline
\multirow{3}{*}{\makecell{+ white noise \\ + foreground \\ residual}}  &  1--10 & 1.0017 & 0.0475 & 0.1413 & 0.1406 & 85.02/100 (0.85) & 84.00 & 70.40 & 1.0117 $\pm$ 0.0482 \\
                 & 11--20 & 0.9917 & 0.0307 & 0.0971 & 0.1017 & 120.99/207 (0.58) & 76.20 & 70.60 & 0.9926 $\pm$ 0.0279 \\
                 & 21--30 & 1.0080 & 0.0272 & 0.0974 & 0.1174 & 115.83/295 (0.39) & 57.80 & 50.80 & 1.0031 $\pm$ 0.0253 \\
\hline 
\end{tabular}}
\label{tab:results-21cmWNFG-APS-mocks}
\end{table*}

\subsection{Tests over N-body mocks} \label{sec:mock-results}

An additional test of the BAO fitting pipeline concerns its application over the set of 500 N-body mocks, generated through a completely different methodology, as described in Section \ref{sec:mocks}. 
We employ the fiducial configuration to analyze the 21\,cm only simulations and the BINGO-like simulations, constructed using the N-body mocks.
We use the same 10 foreground residual maps as before and a fixed beam size $\theta_{\mbox{\sc fwhm}} = 40$ arcmin. 
The summary statistics from these analyses are presented in Tables \ref{tab:results-21cmWNFG-ACF-mocks} and \ref{tab:results-21cmWNFG-APS-mocks}, for ACF and APS, respectively. 

From the 21\,cm only analyses, we find small differences with respect to what is obtained from the {\tt FLASK} mocks, more expressive for the APS estimator, especially the smaller amplitude of $\langle\chi^2\rangle$ and fractions $N_s$ and $N_d$ for all the redshift ranges.
However, such differences can be partially explained by the number of N-body mocks available, one third of the number of {\tt FLASK} mocks. 
In fact, the smaller number of simulations implies, for example, a smaller signal-to-noise ratio from the mean clustering measurements, providing a worse fitting over them (note the slightly large bias on $\alpha^m$ from small and intermediate z-bins for both estimators).
Additionally, for the APS (ACF), the number of $dof$, increases (decreases) with the redshift, making the covariance matrix and its inverse the more biased the higher (smaller) the redshift, explaining the discrepancy in the $\langle \chi^2 \rangle$ and the uncertainty estimates. 
In this sense, we can argue that the slight differences between other summary statistics from N-body and {\tt FLASK} mocks seems to be of statistical origin, although we also need to take into account that the two types of mocks are generated following very different methodologies. 

The impact of introducing noise and foreground residuals to the N-body mocks is very similar to that obtained using the {\tt FLASK} mocks.
The greater impact appears on intermediate and high z-bins, for the same reasons previously discussed (see Sections \ref{sec:results-flask-noise} and \ref{sec:results-flask-foreg}).

\subsection{Wiggle \textit{versus} no-wiggle template fitting} \label{sec:significance}

In order to evaluate the goodness-of-fit of the $\omega(\theta)$ and $C_\ell$ templates to the corresponding measurements from each mock, we compare the minimum $\chi^2$, evaluated at the best-fit $\alpha_{bf}$ value, to that obtained by using a no-BAO template, $\chi^2_{nw}$. 
For this, we construct a template model for each estimator (the same Eqs. (\ref{eq:APS-temp}) and (\ref{eq:ACF-temp})) using $C^{temp}(\ell)$ obtained by imposing $P^{temp}(k) = P^{nw}$, instead of using the parameterization from Eq. (\ref{eq:P_k}). 
The square root of the difference among the two quantities, $S = \sqrt{\Delta \chi^2} = \sqrt{\chi^2 - \chi^2_{nw}(\alpha_{bf})}$, given in term of number of $\sigma$, is commonly used to evaluate the statistical significance of the BAO signal detection \cite[see, e.g.,][]{2017/villaescusa, 2019/camacho, 2021/abbott-des}. 
It means one can say that the BAO model (baryon signature) is preferred by $\Delta \chi^2$ compared to a model with no BAO.  

\begin{figure}[h]
\includegraphics[width=0.95\columnwidth]{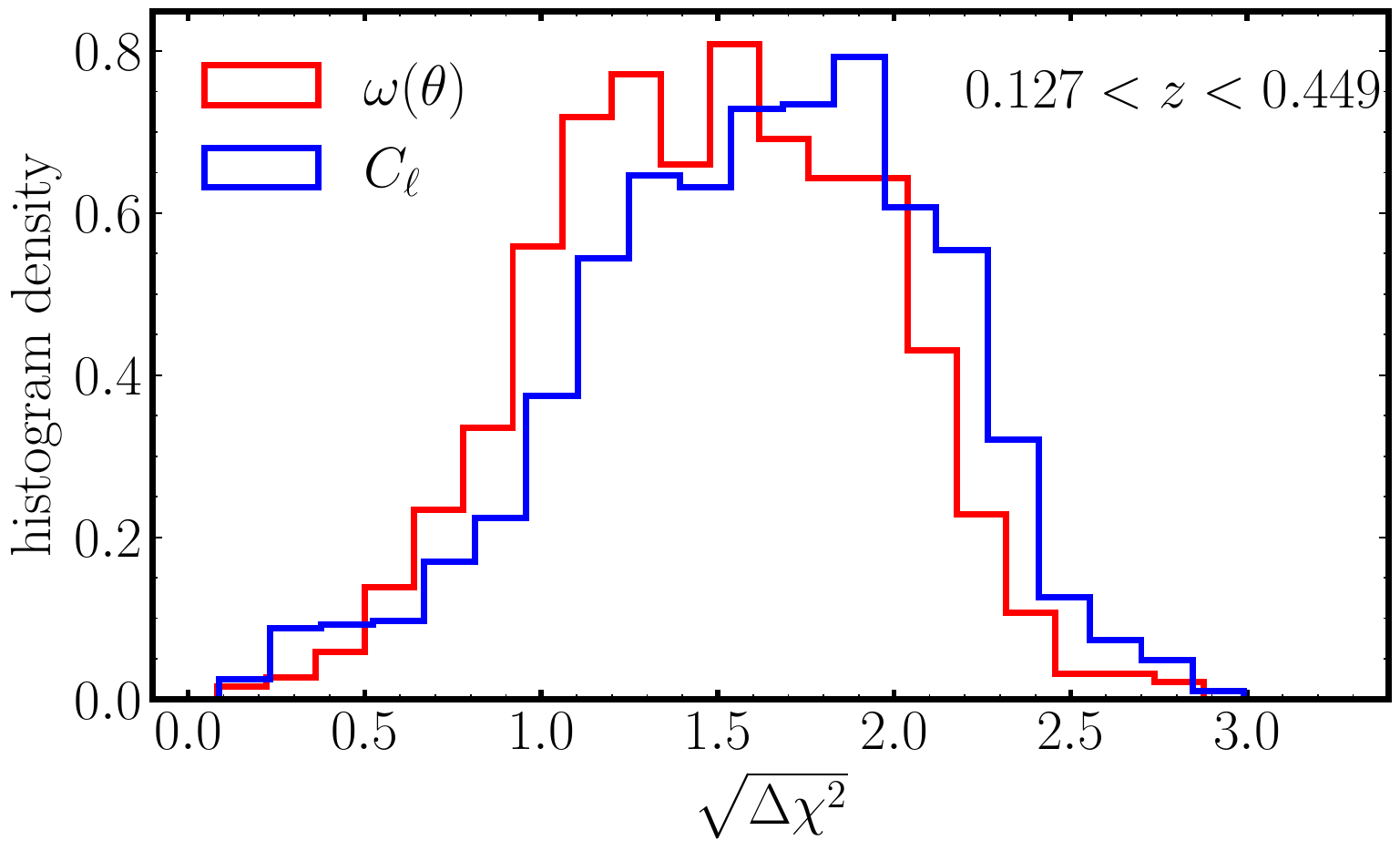}
\caption{Histogram distribution of the significance, $S = \sqrt{\Delta \chi^2}$, of the BAO signal, in terms of number of $\sigma$, estimated for each of the {\tt FLASK} simulations (that is, the corresponding $N_s$ fraction of mocks) considering each clustering estimator. These results are obtained evaluating the BINGO-like simulations combining all the 30 z-bins. 
}
\label{fig:significance}
\end{figure}

To evaluate the effective significance of our BAO detection over BINGO-like simulations we consider the complete BINGO redshift range, fitting all the 30 z-bins together instead of three subsets of them. 
Since the number of dof increases significantly in this case, namely, 256 and 604 for ACF and APS, respectively, an unbiased estimate of the covariance matrix and its inverse requires a larger set of simulations, which directly influences the $\chi^2$ estimate (see discussion in Section \ref{sec:mock-results}). 
For this we rotate six times the original 1500 {\tt FLASK} mocks, similarly to the procedure used to enlarge the number of N-body mocks, as described in section \ref{sec:mocks}, achieving a total of 10500 mocks. 
The clustering statistics measured from the 10500 BINGO-like simulations are used to estimate the covariance matrix, while the $\Delta\chi^2$ quantity is calculated only for the 1500 original mocks. 
The distribution of $\sqrt{\Delta\chi^2}$ from the corresponding $N_s \sim 90$\% and $94$\% of the mocks for ACF and APS, with average significance of $\langle S \rangle = 1.48\sigma$ and $1.61\sigma$, respectively, are presented in Fig. \ref{fig:significance}. 
From the $N_s$ fraction of mocks, we find $\sim 14$\% and $\sim 24$\% of them with $S = \sqrt{\Delta\chi^2} > 2\sigma$, for ACF and APS, respectively. 
Confirming the slightly better performance of the APS over the ACF estimator, the former one seems to provide BAO detection with higher significance than the last one. 
Also, although the $\sqrt{\Delta \chi^2}$ distribution is shown only for the BINGO-like simulations, notice that it is impacted by systematic effects similarly to other summary statistics. 
Compared to 21 cm only case, $\sqrt{\Delta \chi^2}$ is significantly shifted to smaller values due to the inclusion of thermal noise, while the contribution of foreground residual has a negligible impact.

\section{Conclusions} \label{sec:conclusions}

The BAO feature measured from galaxy distribution is currently recognized as one of the most important cosmological probes. 
Its dependence with the evolution and the components of the Universe makes it a powerful tool to investigate the physical nature of dark energy. 
Since its first detection, the BAO scale has been measured using different matter tracers in several redshift ranges, and now the radio 21\,cm IM experiments, such as BINGO, plans to provide a complementary route to study BAO. 

Here we present the assessment of the BINGO telescope potential, in its Phase 1 operation, to detect the transversal BAO signal. 
For this we developed a template fitting pipeline with two clustering estimators, $\omega(\theta)$ and $C_\ell$, to extract the BAO information from two different sets of simulations, {\tt FLASK} and N-body mocks. 
Our analyses account for the BINGO sky coverage, beam effect and evaluate the impact of including thermal noise and foreground residual. 
The fitting procedure to measure the shift parameter, $\alpha$, from the mocks is performed over three sets of consecutive z-bins using a maximum likelihood estimator. 
Most of our analyses are performed employing the fiducial configuration (aspects (a) and (b) defined in Section \ref{sec:results}), optimized to be applied to all redshift ranges. Our main results can be summarized as follows: 

\begin{itemize}
    \setlength\itemsep{0.15cm}
    \item We found that both estimators perform better at higher redshifts, but show different sensitivities to specific redshift ranges. 
    The bias amplitude in $\langle \alpha \rangle$, the $\sigma_{68}$ uncertainty, and the selection and detection fractions, $N_s$ and $N_d$, suggest that the APS slightly outperforms the ACF estimator, although the ACF presents slightly smaller $\sigma_{68}$ at lower z-bins. 
    Also, the APS leads to greater average significance, $\langle S \rangle$, than the ACF.
    This behavior is observed analyzing 21\,cm only simulations and remains after including contaminant signals and redshift dependent beam size (see Tables \ref{tab:results-21cmWNFG-ACF} to \ref{tab:results-21cmWNFG-APS-mocks}). 
    
    \item Accounting for the impact of including thermal noise and foreground residual one at a time, we found that the former one has the greater impact in both clustering estimators, affecting results more significantly the greater the redshift. 
    It enlarges significantly the $\sigma_{68}$ error, in particular for the APS estimator, reaching 122\% at higher z-bins. 
    The reason for that is that the BAO wiggles appear at higher multipoles (smaller angular scales) the higher the redshift, scales at which the thermal noise starts dominating. 
    The inclusion of the foreground residual, on the other hand, does not seem to have a significant impact in our results.
    
    \item Robustness tests of the fiducial configuration, considering different $\Delta \theta$ and $\Delta \ell$ binning and alternative template models ($q$ indexes), suggest that 
    a more appropriate choice would be to select an "optimal configuration" for each of the three redshift ranges.
    
    \item Using the fiducial configuration, the BAO fitting pipeline applied to 500 N-body mocks (Tables \ref{tab:results-21cmWNFG-ACF-mocks} and \ref{tab:results-21cmWNFG-APS-mocks}) show a few aspects differing from those analyzing {\tt FLASK} mocks, caused mainly by the smaller number of N-body mocks. 
    
    \item We also evaluate how preferred a BAO model is when compared to a fit of a non-BAO model, and find an average significance for the BAO detection of $\sim 1.48\sigma$, for ACF, and $\sim 1.48\sigma$, for APS, combining all 30 z-bins. A fraction of 14\% and 24\% of the {\tt FLASK} mocks, for ACF and APS, respectively, provide a $> 2\sigma$ BAO detection.
\end{itemize}

It is worth emphasizing that some aspects of the analysis presented here could benefit from the inclusion of more realistic choices of instrumental issues, such as the structure of the beam and instrumental effects like $1/f$ noise and polarization leakage, or even aspects of the methodology that can be improved and deserve further study. 
Among them we can mention the test of alternative component separation procedures, a different number of z-bins for the BAO fitting \citep[see results from the BINGO paper,][]{2022/mericia_BINGO-component_separation_II}, and a fitting method alternative to MLE.

Moreover, it is well known \citep{2007/eisenstein} that the non-linear gravitational evolution of the Universe smears out the acoustic signature by inducing a damping and broadening on the BAO peak, besides shifting its position so that, the lower the redshift, the more difficult is to achieve an accurate measurement of the BAO features. 
Since BINGO will survey a large volume of the Universe at relative low redshifts, we also expect to improve our BAO fitting results, in particular at lower redshifts, by considering a reconstruction procedure \citep{2017/obuljen}. 

In summary, we conclude that intermediate and higher redshift intervals are the most promising in measuring the BAO scale, with a probability of detection of more than $\sim 80\%$ ($\sim 70\%$) with the APS (ACF) estimator. 
Such numbers are obtained using a fixed fiducial configuration and can be significantly improved by the choice of an "optimal configuration" for each redshift interval.
Indeed, although showing that systematic effects have a non-negligible impact on $\alpha$ estimates, our analyses of a semi-realistic scenario (BINGO-like mocks) confirm that BINGO should be able to successfully detect the BAO signal in radio frequencies.

\begin{acknowledgements}
The BINGO project is supported by São Paulo Research Foundation (FAPESP) grant 2014/07885-0.
C.P.N. would like to thank Edilson de Carvalho, Armando Bernui, Henrique Xavier, and Hugo Camacho for very enlightening and useful discussions. C.P.N. also acknowledges FAPESP for financial support through grant 2019/06040-0. 
J.Z acknowledges support from the Ministry of Science and Technology of China (grant Nos. 2020SKA0110102).
R.G.L. thanks CAPES (process 88881.162206/2017-01) and the Alexander von Humboldt Foundation for the financial support. 
L.S. is supported by the National Key R\&D Program of China (2020YFC2201600).
A.A.C acknowledges financial support from the National Natural Science Foundation of China (grant 12175192).
L.B., F.A.B, A.R.Q., and M.V.S. acknowledge PRONEX/CNPq/FAPESQ-PB (Grant no. 165/2018).
E.F.: The Kavli IPMU is supported by World Premier International Research Center Initiative (WPI), MEXT, Japan.
We thank an anonymous referee for her/his very insightful comments.
This research made use of {\tt astropy} \citep{2018/astropy}, {\tt healpy} \citep{2019/healpy}, {\tt numpy} \citep{2011/numpy}, {\tt scipy} \citep{2020/scipy} and {\tt matplotlib} \citep{2007/matplotlib}.
\end{acknowledgements}

%
%

\bibliographystyle{aa} 
\bibliography{BINGOpaperVIII} 

\end{document}